\documentclass[aps,prd,nofootinbib,reprint,superscriptaddress]{revtex4-2}
\usepackage{hyperref}
\usepackage{graphicx}

\usepackage{xcolor}
\usepackage{subcaption}
\usepackage{csquotes}
\usepackage{subcaption}
\usepackage[percent]{overpic}
\usepackage{dcolumn}
\usepackage[normalem]{ulem}
\usepackage{booktabs}

\begin{document}
\preprint{APS/123-QED}
\title{Probing the chiral magnetic effect via transverse spherocity event classification in relativistic heavy-ion collisions}
	\author{Somdeep Dey}
    \affiliation{School of Physics, University of Hyderabad, Gachibowli, Hyderabad, India 500046}

	\author{Abhisek Saha}
	\affiliation{School of Physics, Peking University, Beijing 100871, China}
	\affiliation{Center for High Energy Physics, Peking University, Beijing 100871, China}

\begin{abstract}
    We present the first study of the Chiral Magnetic Effect (CME) using transverse spherocity as an event-shape classifier in Pb+Pb collisions at $\sqrt{s_{NN}} = 5.02$ TeV, simulated with the A Multi-Phase Transport (AMPT) model with a realistic CME implementation. Transverse spherocity separates events into jetty and isotropic topologies based on the geometric distribution of transverse momentum. Unlike traditional event shape engineering methods, which use the flow vector as an event classifier that is itself contaminated by the very backgrounds it is intended to suppress, spherocity provides a cleaner, geometry-driven classification that avoids this circular limitation. CME inclusion shifts the spherocity distribution toward more isotropic events, confirming its sensitivity to CME-induced charge separation. The charge-dependent azimuthal correlator $\Delta\gamma$ and correlated background coupled with elliptic flow are consistently higher in jetty events. The scaled ratio $\Delta\gamma/v_2$ shows enhanced values for isotropic events, confirming effective background suppression after elliptic flow scaling. 
    Our results demonstrate that isotropic event selection via transverse spherocity provides a cleaner and more reliable environment for CME searches by simultaneously suppressing flow-driven and resonance-decay backgrounds, making it a powerful complementary method to existing flow-vector-based methods.
\end{abstract}


\maketitle
\section{Introduction} 			\label{sec:intro} 
Relativistic heavy-ion collisions create extreme conditions of temperature and energy density that allow us to study the fundamental properties of matter governed by Quantum Chromodynamics (QCD) \cite{Kharzeev:2007jp, Bali:2011qj, Bzdak:2019pkr, An:2021wof}. In these collisions, ordinary matter made of confined quarks and gluons melts into a new state called the quark-gluon plasma (QGP), where quarks move freely, and chiral symmetry is restored \cite{STAR:2014uiw}. Experiments at the Relativistic Heavy Ion Collider (RHIC) and the Large Hadron Collider (LHC) have confirmed that this hot and dense matter behaves almost like a perfect liquid, showing strong collective behavior across a wide range of collision energies and centralities\cite{Xu:2025plf,  Shuryak:2014zxa, Kharzeev:2013jha, Jie:2018xvh}. Apart from these bulk properties, these collisions offer a way to probe subtle, symmetry-driven phenomena inherent to the QCD vacuum. The Chiral Magnetic Effect (CME) is one such phenomenon that links the topological aspects of QCD with observable electric charge separation \cite{Yuan:2024wpz, Fukushima:2008xe, Kharzeev:2015znc, Kharzeev:2013ffa, Fukushima:2010fe, Buividovich:2009wi}.\\

The CME is driven by two short-lived phenomena that occur in non-central heavy-ion collisions. The ultra-relativistic motion of the colliding positively charged nuclei creates a strong but short-lived magnetic field, estimated to reach magnitudes of up to $10^{14}-10^{15}$T shortly after the collision \cite{Adhikari:2024bfa, Bali:2011qj, Bzdak:2011yy, Kharzeev:2013jha}. This field is oriented, on average, perpendicular to the reaction plane (RP), which is defined by the beam axis and the impact parameter direction of the colliding nuclei. At the same time, the QGP is expected to contain local P- and CP-odd domains characterized by nontrivial topological gluon field configurations \cite{Shou:2014zsa, Zhao:2019hta, Kharzeev:2007jp, Kharzeev:2009fn, Voloshin:2004vk}. These domains can induce a chirality imbalance, resulting in a difference in the number of left- and right-handed quarks, which manifests as a non-zero axial chemical potential. In the presence of a strong magnetic field, such a chirality imbalance can generate a macroscopic electric current along the field direction, leading to the separation of electric charges with respect to RP \cite{Buividovich:2008wf, Asakawa:2010bu}. This CME current is a direct transport signature of QCD topology and chiral symmetry restoration in the QGP \cite{Schlichting:2010qia, Jie:2018xvh, Kharzeev:2024zzm, Kharzeev:2004ey, Zhao:2019hta, Kharzeev:2013ffa, Kharzeev:2007jp, Feng:2025yte, Grieninger:2025spw}. Observing this charge-separation experimentally would provide direct evidence for local parity violation in strong interactions and reveal important properties of the QCD vacuum under extreme conditions.\\ 

Extensive experimental and theoretical efforts have been devoted to searching for CME-induced charge separation in relativistic heavy-ion collisions.
Experiments at RHIC and LHC have measured non-zero charge-dependent azimuthal correlations ($\Delta\gamma$) that are qualitatively consistent with expectations of the CME \cite{STAR:2009wot, STAR:2009tro, STAR:2025uxv, Feng:2025yte, STAR:2021mii, STAR:2014uiw, Koch:2016pzl, Jie:2018xvh, Zhao:2019hta}. However, a conclusive isolation of the CME signal has remained elusive \cite{STAR:2009wot, STAR:2009tro, ALICE:2012nhw, Kharzeev:2015znc, Zhao:2018ixy, Zhao:2019hta}. Most recently, the dedicated isobar run at RHIC, which compared collisions of $^{96}_{44}$Ru and $^{96}_{40}$Zr to control the magnetic field strength while keeping the background conditions similar, did not yield a definitive CME observation \cite{STAR:2021mii}. This highlights how difficult it is to isolate the CME in practice.\\

The core difficulty lies in the background. Several processes can produce very similar charge-dependent azimuthal correlations that are parity-even and unrelated to topological fluctuations. These backgrounds arise from correlations between daughter particles from a resonance decay, local charge conservation effects coupled with collective flow, and from correlations between particles emitted in jet fragmentation \cite{Wang:2009kd}. Crucially, all of these background contributions are strongly tied to the elliptic flow coefficient $v_2$, which quantifies how anisotropic the particle emission is in the transverse plane \cite{Voloshin:2004vk, Wang:2009kd, Liao:2010nv, Bzdak:2010fd, Pratt:2010zn, STAR:2015wza}. Because both the CME signal and the background correlations are measured relative to the same reaction plane, and because the backgrounds scale approximately with $v_2$, disentangling a genuine CME signal from these flow-driven correlations is extremely challenging. What is needed, therefore, is an event classifier that can separate events by their underlying geometry without relying on $v_2$ itself as the primary handle.


Many techniques have been developed since then to eliminate or mitigate these backgrounds \cite{STAR:2013zgu, STAR:2014uiw, CMS:2016wfo, STAR:2019xzd, ALICE:2020siw, Wang:2025dfz, Deng:2016knn}, including innovative observables \cite{ALICE:2017sss, CMS:2017lrw, Voloshin:2018qsm}. One widely used approach is event shape engineering (ESE) \cite{ALICE:2017sss, CMS:2017lrw, Schukraft:2012ah}. In this method, the events are grouped into classes of different $v_2$ values depending on the dynamical and statistical fluctuations of $v_2$ \cite{ALICE:2026cvp, Li:2025ohn, ALICE:2017sss, Li:2024gdz}.
By comparing events with different flow strengths, they examine how backgrounds change and isolate the CME contribution. However, this approach has inherent limitations because ESE uses $v_2$ as the classifier, while $v_2$ itself is already entangled with the backgrounds it is intended to suppress. A classifier that is more directly connected to the initial collision geometry rather than to the final-state flow magnitude could provide a cleaner and more powerful separation.\\

Transverse spherocity ($S_0$) is precisely such a classifier. It is a novel event-shape variable that characterizes events based on the geometrical distribution of transverse momentum in an event \cite{Mallick:2020ium}. It characterizes events based on how isotropic or collimated the transverse momentum distribution of particles is in a given event. It is a powerful tool for separating jetty events, in which particles are emitted in a narrow cone, from isotropic events, in which particles spread more uniformly. Crucially, isotropic events are expected to have lower jet contributions and reduced flow-driven backgrounds, making them better candidates for CME searches. Recent results from LHC experiments confirm that spherocity is a useful event classifier in both p+p \cite{Rathore:2025wgx, Behera:2025ycg, CMS:2025sws, ALICE:2023bga, ALICE:2019dfi} and $^{208}_{\ 82}$Pb+$^{208}_{\ 82}$Pb collisions \cite{Mallick:2020ium, Tripathy:2026atg, Prasad:2025dbk, Prasad:2025ezg, Tripathy:2025npe}. Preliminary measurements by the ALICE collaboration at the LHC further suggest that spherocity-based event selection can enhance or suppress anisotropic flow and jet-related backgrounds by separating event types, making it a promising tool to address the limitations of existing methods in CME searches \cite{Mallick:2020ium, Pucillo:2025zof, ALICE:2023bga, Deb:2020ige}.\\

In this study, we present the first comprehensive analysis of the CME using transverse spherocity as an event-shape classifier. We incorporate a realistic CME implementation into the AMPT (A Multi-Phase Transport) framework and analyze events based on their spherocity values. AMPT is well-suited for this study because it incorporates both partonic and hadronic interactions and allows for a controlled implementation of the CME. Our analysis has three main goals: first, to show that spherocity effectively separates event topologies and is sensitive to CME implementation; second, to study how the $\Delta\gamma$ correlator and its key backgrounds, such as elliptic flow and resonance decays, behave across different spherocity classes; and third, to identify the optimal spherocity selection that maximizes sensitivity to a potential CME signal by suppressing conventional backgrounds.\\

This paper is organized as follows. Section \ref{sec:method} describes the CME observables, the background mechanisms, the transverse spherocity variable, and the AMPT model with CME implementation. Section \ref{sec:results} presents our findings, including how spherocity distributions change with CME and how key backgrounds behave across spherocity classes. Following that, we analyze how the charge-dependent correlator $\Delta\gamma$ and the scaled correlator $\Delta \gamma/v_2$ respond to different spherocity selections. Throughout this section, we explore how different spherocity cuts affect the separation between jetty and isotropic events. Finally, Section \ref{sec:conclusion} summarizes our main results and discusses their implications for future CME searches in heavy-ion experiments.


\section{Methodology}
\label{sec:method}
\subsection{CME observable and correlated background signal}
 Experimentally, the CME is expected to manifest as a charge separation perpendicular to the reaction plane, spanned by the impact parameter and the beam momenta in a collision. In the study of the CME-induced charge separation as well as other collective motions in the QGP, the azimuthal distribution of produced particles is often expressed with the Fourier expansion for given transverse
momentum ($p_T)$) and pseudorapidity ($\eta$) in an event \cite{voloshin1996flow,Voloshin:2004vk}:
\begin{equation}
    \frac{dN_{\pm}}{d\phi} \propto 1 \pm 2a_1\sin\phi + 2v_2\cos 2\phi + \cdot\cdot\cdot,
\end{equation}
where $\phi$ is the azimuthal angle of the particle momentum vector with respect to the RP and the subscript $'\pm'$ indicates particle charge sign. The elliptic anisotropy $v_2$  is the leading modulation in particle distributions
produced in relativistic heavy ion collisions. Although the $a_1$ coefficient can quantify the CME-induced charge separation, the CME signal fluctuates event-by-event and makes the $\langle\sin(\phi)\rangle$ zero \cite{Bzdak:2011yy}. Thus, it cannot be directly observed from single-particle distributions. Instead, multi-particle correlation observables sensitive to charge-dependent azimuthal correlations relative to the reaction plane are required to measure $a_{1,\pm}$ fluctuations across the RP, such as the $\gamma$ correlators \cite{Voloshin:2004vk}, the $R$ correlators \cite{PhysRevC.97.061901, Yuan:2024wpz, Ajitanand:2010rc}, and the signed balance functions \cite{Tang:2019pbl, Lin:2020jcp}. One of the most widely used observables for this purpose is the charge-dependent three-particle correlator, commonly referred to as the $\gamma$ correlator, defined as\cite{Huang:2017pzx, Bzdak:2012ia}
\begin{equation}
\gamma_{\alpha\beta} =
\left\langle \cos(\phi_\alpha + \phi_\beta - 2\Psi_{\mathrm{RP}}) \right\rangle ,
\label{eq:gamma}
\end{equation}
where the averaging is carried out over all charge particle pairs $(\alpha, \beta)$ within each event and across all events.

To suppress charge-independent correlations such as effects from global momentum conservation and enhance sensitivity to CME-induced charge separation, the difference between opposite-sign (OS) and same-sign (SS) correlations is commonly studied \cite{STAR:2009wot, STAR:2009tro},
\begin{equation}
\Delta\gamma = \gamma_{\mathrm{OS}} - \gamma_{\mathrm{SS}} ,
\label{eq:deltagamma}
\end{equation}
The backgrounds result from correlations between particles from the same jet or back-to-back dijet, or between daughter particles from a resonance decay, coupled with elliptic flow anisotropies, can be expressed as a two-particle correlation \cite{Voloshin:2004vk},
\begin{equation}
    \Delta \gamma_{res}  = \langle\cos(\phi_\alpha + \phi_\beta - 2\phi_{res})\rangle v_{2,res}
    \label{eq:deltagammares}
\end{equation}
where the subscript “res” represents correlated
two-particle clusters, like resonances and jets. The $v_2$ of these background sources is defined as,
\begin{equation}
    v_{2,res} = \langle \cos\left[2(\phi_{res} - \Psi_2)\right]\rangle
    \label{eq:v2res}
\end{equation}
We demonstrate how these backgrounds change for different sphericity selections in Sec-\ref{sec:resB}.

\subsection{Introduction of CME in AMPT model}
Our analysis uses the AMPT model to study how particles behave in heavy-ion collisions \cite{Lin:2004en, Lin:2000cx}.
We simulate $5 \times 10^{5}$ minimum-bias events in Pb+Pb collisions at $\sqrt{s_{\mathrm{NN}}} = 5.02$ TeV using the string-melting version of the model. The AMPT framework begins by initializing collisions with the HIJING model, which generates the spatial and momentum distributions of minijet partons and soft string excitations \cite{Wang:1991hta}. Subsequent stages include partonic scattering via Zhang's Parton Cascade (ZPC), hadronization through a coalescence mechanism, and a final hadronic rescattering phase. This version of the model has been shown to successfully describe key observables in heavy-ion collisions, including particle spectra and anisotropic flow. 
For the present analysis, we define the reaction plane to lie in the $X$-$Z$ plane, aligning the average magnetic field direction along the negative $y$-axis.\\

The standard AMPT model does not include mechanisms to generate the chiral magnetic effect. To incorporate the CME, we follow a well-established prescription where an initial charge separation is imposed at the partonic stage. This implementation, detailed in our previous work \cite{Dey:2026wve, Dey:2025pxi}, is based on the methodology introduced in Ref. \cite{Ma:2011uma}. The procedure selectively exchanges the $p_y$ momentum components between a fraction of downward-moving quarks and upward-moving antiquarks of the same flavor. This selection is defined by the charge separation fraction $f$. This exchange induces a net charge separation along the magnetic field direction while rigorously conserving the total momentum of the system. By varying the charge separation fraction $f$ of affected partons, we can control the strength of the introduced CME signal, enabling a systematic study of its impact on spherocity-dependent observables.

\subsection{Transverse Spherocity as an event classifier}
Transverse spherocity provides a quantitative measure of the event topology by characterizing whether the transverse momentum flow is distributed isotropically or is dominated by jet-like structures. It is defined as \cite{Mallick:2020ium},

\begin{equation}
S_0 = \frac{\pi^2}{4}
\min_{\hat{n}}
\left(
\frac{\sum_i \left| \vec{p}_{T,i} \times \hat{n} \right|}
{\sum_i p_{T,i}}\right)^2 ,
\label{eq:spherocity}
\end{equation}

where $\vec{p}_{T, i}$ is the transverse momentum vector of the $i$-th particle and $\hat{n}$ is a unit vector in the transverse plane that minimizes the expression. The normalization factor ensures that the spherocity values are within the range $0 \leq S_0 \leq 1$. Values close to zero mean the event is jetty. These jetty events have high-momentum particles aligned in a specific direction. On the other hand, values close to $1$ indicate the event is isotropic. In isotropic events, particles are emitted uniformly in all directions within the transverse plane. This intrinsic sensitivity allows spherocity to discriminate between events rich in jet-like correlations and those dominated by the collective bulk medium. Consequently, it offers a novel and powerful tool to investigate and suppress background sources that may mimic the chiral magnetic effect.\\

\section{Results}
\label{sec:results}
\subsection{Spherocity Distribution in the Presence of Chiral Magnetic Effect}  
\label{sec:resA}

We analyze $^{208}_{\ 82}$Pb+$^{208}_{\ 82}$Pb collisions at $\sqrt{s_{NN}} = 5.02$ TeV for $(30$-$50)\%$ and $(50$-$70)\%$ centrality class. The spherocity is calculated within the pseudo-rapidity range $|\eta| < 0.8$, with an additional requirement of at least $10$ charged particles having transverse momentum $p_T > 0.15$ GeV/$c$, to closely match the acceptance conditions of the ALICE detector at the LHC \cite{Mallick:2020ium}.
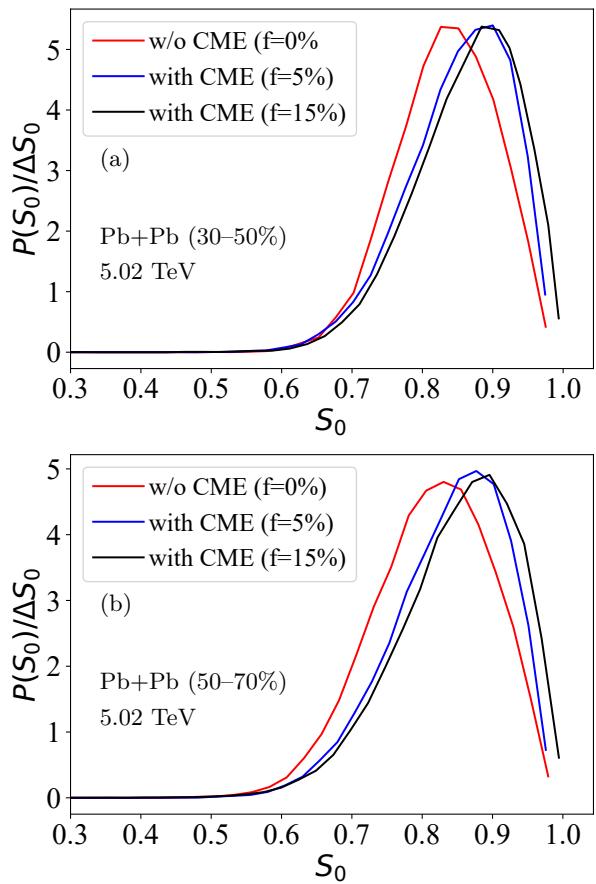
\begin{figure}[h!]
\centering

\begin{overpic}[scale=0.5]{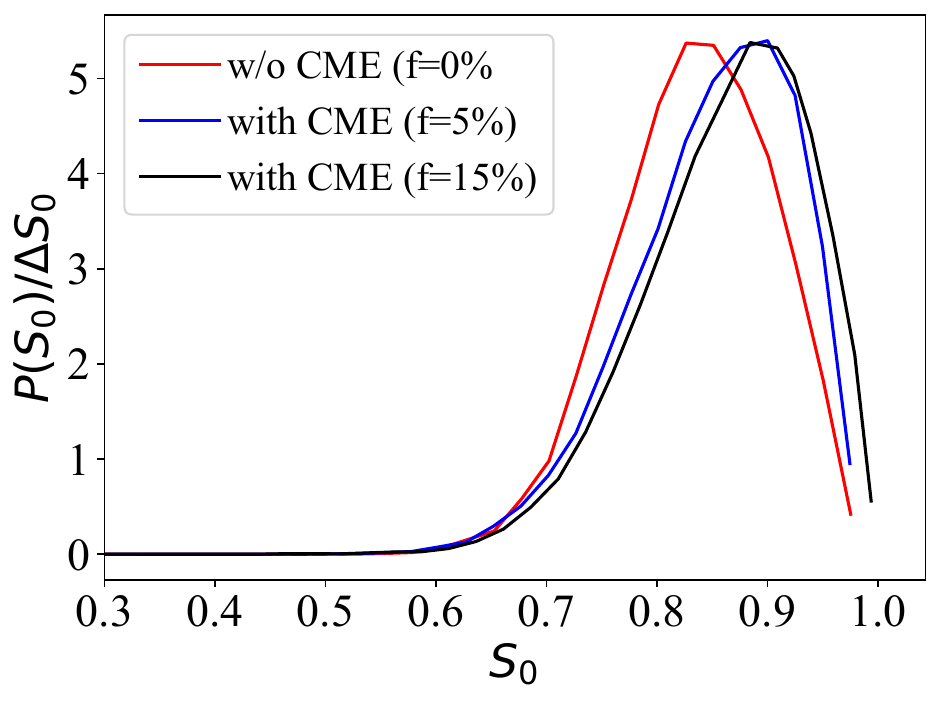}
\put(16,47){(a)}
\put(16,34){Pb+Pb (30–50\%)}
\put(16,28){5.02 TeV}
\end{overpic}

\begin{overpic}[scale=0.5]{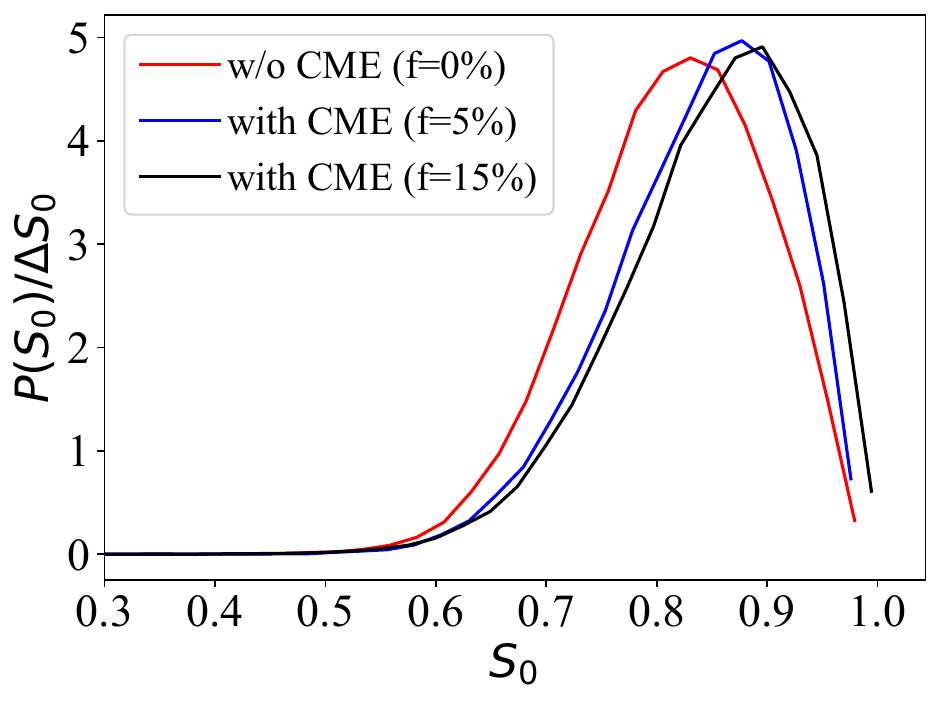}
\put(16,47){(b)}
\put(16,34){Pb+Pb (50–70\%)}
\put(16,28){5.02 TeV}
\end{overpic}
\caption{Transverse spherocity distributions for Pb+Pb collisions at $\sqrt{s_{NN}} = 5.02$ TeV in the (a) $(30$-$50)\%$ and (b) $(50$-$70)\%$ centrality classes. Results are presented for the AMPT model both without and with the CME implementation}
\label{fig:sph}
\end{figure}
Figure \ref{fig:sph} presents the transverse spherocity distributions obtained before and after the implementation of CME in the AMPT model, for the two centrality intervals considered in this study. A systematic shift of the distributions towards higher $S_0$ values is observed in the presence of CME. 
This shift corresponds to a relative enhancement of isotropic events ($S_0\rightarrow$1) and a suppression of jetty configurations ($S_0\rightarrow$0) in the final-state particle distribution. The effect is visibly similar in both centrality classes, with a wider distribution in peripheral collisions than in the mid-central (30–50)\% class.\\

 This systematic enhancement towards higher spherocity can be understood from the mechanism of CME implementation. The CME implementation involves exchanging the $p_y$ components between downward-moving quarks and upward-moving antiquarks of the same flavor. The net result of this specific, momentum-conserving exchange is not a simple, uniform dispersion of momentum in all direction. Instead, it enhances the existing charge-dependent $p_y$ separation. 
 This increases the charge separation along the y-axis (the magnetic field direction).
 Spherocity finds the axis $\hat{n}$ that minimizes the sum of projected transverse momentum magnitudes. An event is "jetty" ($S_0$ low) if most transverse momentum is aligned along a single axis (the jet direction). It is "isotropic" ($S_0$ high) if momentum is evenly distributed in the transverse plane. The CME implementation does not create a new jet axis. However, it systematically modifies the $p_y$ components of a subset of partons relative to the global reaction plane. This is a collective, global effect tied to the reaction plane ($\Psi_{RP}$, here along the x-axis). For many particles, this introduces an additional momentum component perpendicular to the dominant jet axis (which is random relative to the reaction plane). Adding a momentum component orthogonal to any potential jet axis tends to defocus the event's momentum flow, making it less collimated and more dispersed in azimuth. This dispersion reduces the ability to find a single axis $\hat{n}$ that captures most of the event's momentum, thereby increasing the calculated $S_0$.\\

 Moreover, in the AMPT model, the hardest particles often come from jet fragmentation or initial hard scatterings. The CME-induced $p_y$ exchange acts on partons early in the evolution. If a parton that would have become a high-$p_T$ jet fragment has its $p_y$ significantly altered, it can disrupt the tight, back-to-back angular correlation of the jet cone. This "smearing" of jet constituents weakens the collimated momentum flow that defines a jetty event. The event consequently appears less jet-like and more isotropic to the spherocity.\\

 On the other hand, the spherocity distributions remain unchanged across different values of the separation fraction $f$. Increasing $f$  means increasing the number of partons directly affected by the CME prescription. This affects the azimuthal correlations of the finally produced particles and thus increases the CME signal, as observed in previous studies \cite{Dey:2026wve, Dey:2025pxi, Ma:2011uma}. Consequently, the CME prescription does not add net transverse momentum; it exchanges $p_y$  between selected quarks and antiquarks. For a given quark–antiquark pair that swaps $p_y$, the vector sum of their transverse momenta is conserved. Therefore, the overall momentum configuration, and hence the orientations relative to the optimal axis, is perturbed in a way that may not alter the minimized ratio in Eq. \ref{eq:spherocity} significantly when $f$ increases beyond a certain level.\\

 Based on the observed shift in the spherocity distribution, we can leverage transverse spherocity as a powerful discriminant to isolate a potential CME signal from dominant backgrounds. Our results demonstrate that the inclusion of a CME-like charge separation systematically biases events toward a more isotropic topology ($S_0\rightarrow$1). 
 This provides the foundation for a robust analysis strategy. By categorizing events into isotropic ($S_0\approx$1) and jetty ($S_0 \approx 0$) classes based on their measured spherocity, we can create two distinct groups. Isotropic events are enriched with the topological signature of the CME and its associated collective azimuthal perturbation, while jetty events are dominated by hard processes and flow-related correlations. This classification allows for a differential study of CME-sensitive observables, such as the $\Delta \gamma$ correlator. Crucially, by examining these observables separately in isotropic and jetty classes, and particularly by investigating their scaling with established background proxies like elliptic flow ($v_2$), we can directly test whether the signal behaves as expected for a CME-driven effect or for flow-coupled or jet-related backgrounds.


\subsection{Understanding background in CME studies in Spherocity-Selected Events}
\label{sec:resB}
\subsubsection{Anisotropic Flow}

\begin{figure}[h!]
\centering

\begin{overpic}[scale=0.5]{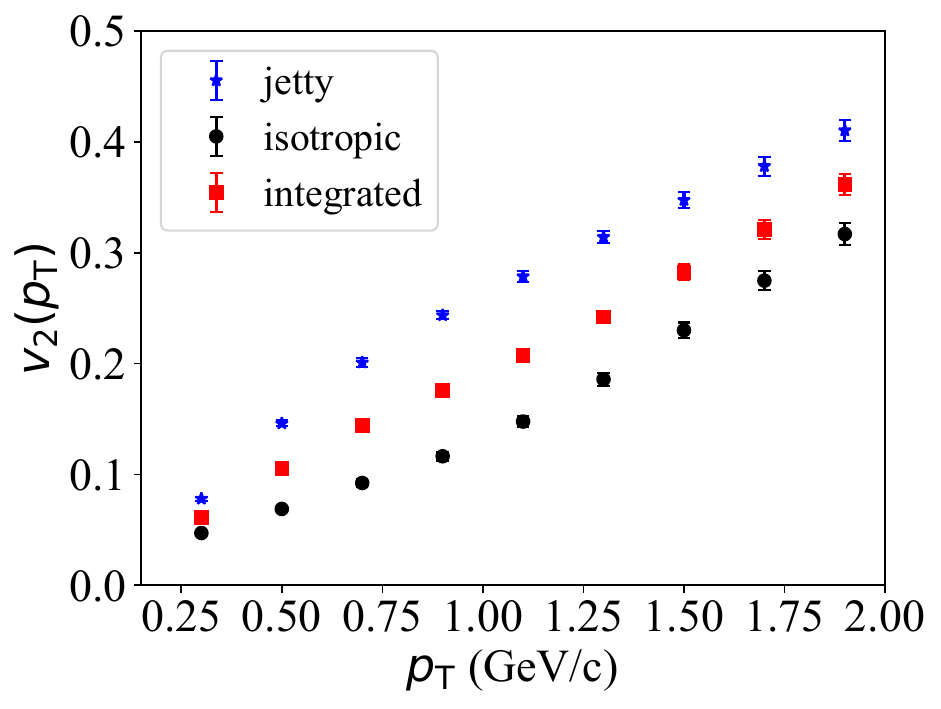}
\put(20,42){\textbf{(a)}}
\put(50,66){Spherocity cut}
\put(52,60){(70\%-30\%)}
\put(60,26){Pb+Pb (30-50\%)}
\put(63,20){5.02 TeV}
\end{overpic}
\begin{overpic}[scale=0.5]{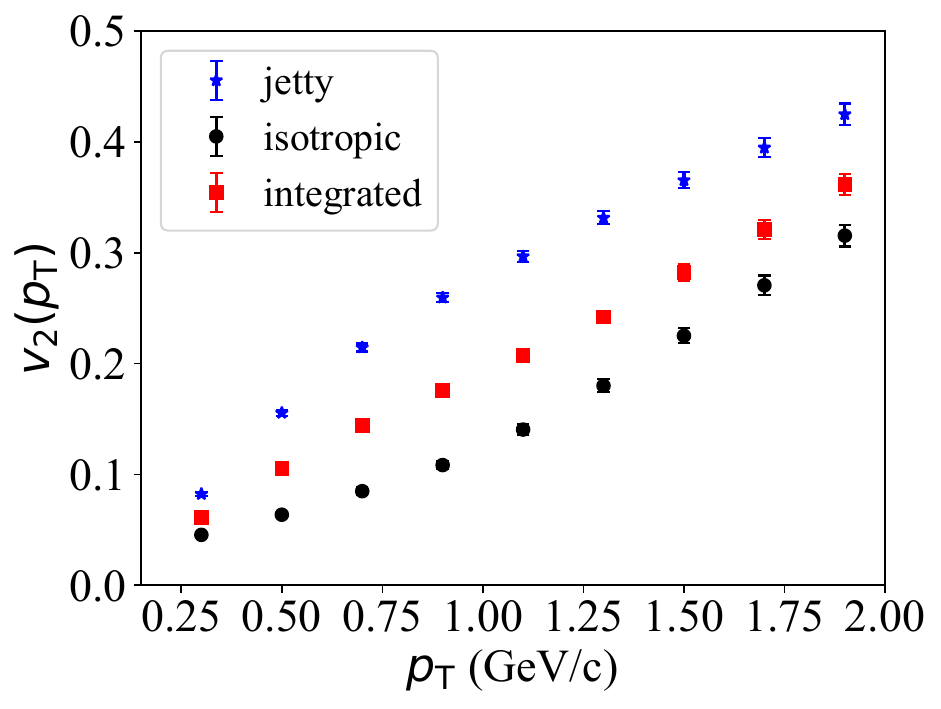}
\put(20,42){\textbf{(b)}}
\put(50,66){Spherocity cut}
\put(52,60){(80\%-20\%)}
\end{overpic}

\vspace{0.5cm}

\begin{overpic}[scale=0.5]{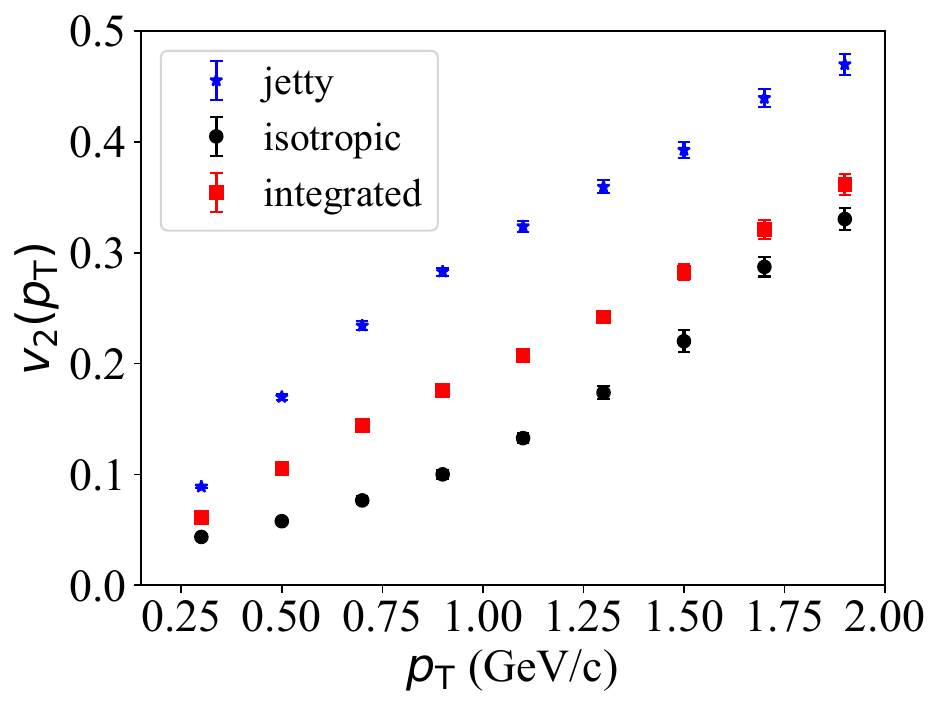}
\put(20,42){\textbf{(c)}}
\put(50,66){Spherocity cut}
\put(52,60){(90\%-10\%)}
\end{overpic}

\caption{Elliptic flow coefficient $v_{2}$ as a function of transverse momentum $p_{T}$ in Pb+Pb collisions at $\sqrt{s_{NN}} = 5.02$ TeV for different transverse spherocity selections (a) $70\%$-$30\%$, (b) $80\%$-$20\%$  and (c) $90\%$-$10\%$}
\label{fig:v2}
\end{figure}
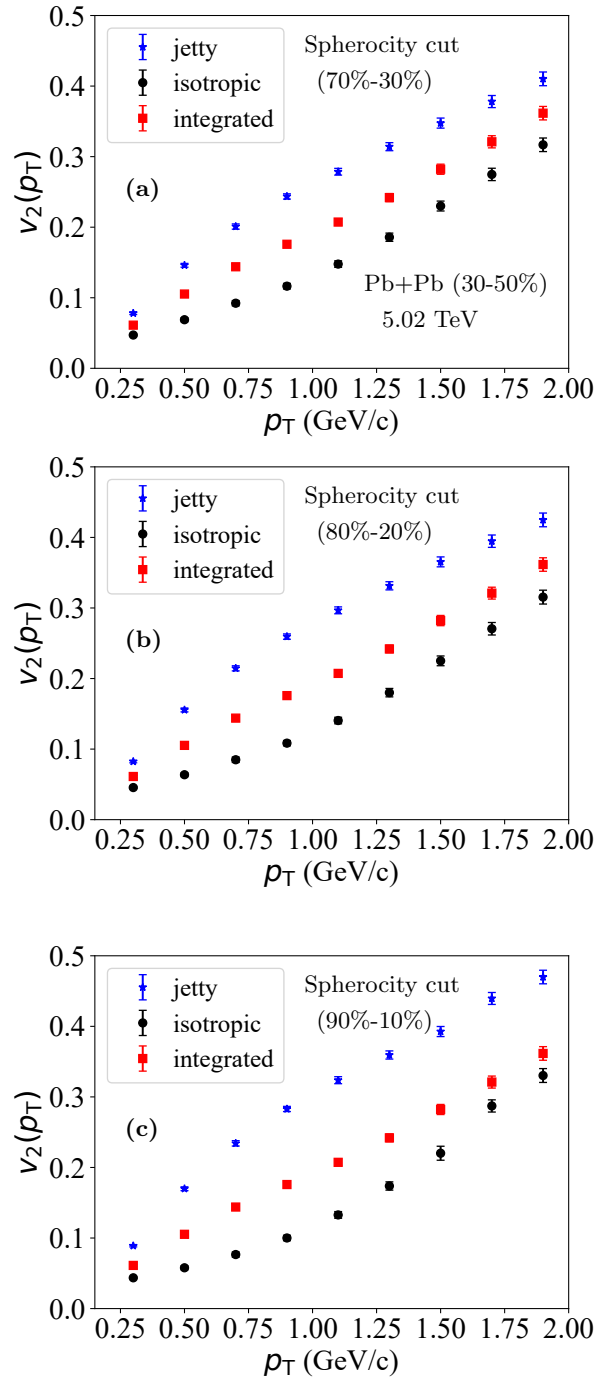

Anisotropic flow is a key manifestation of collective behavior in heavy-ion collisions, arising from the transformation of initial spatial anisotropy in the overlap zone into a momentum-space anisotropy in the final-state particle distributions \cite{ALICE:2016ccg}.  
In the context of CME searches, the second order flow coefficient $v_2$ (defined in Eq. \ref{eq:v2res} for resonance particles)  is of paramount importance because it is intrinsically linked to the dominant background mechanisms, such as local charge conservation coupled with collective expansion, that can produce a non-zero $\Delta\gamma$ correlator indistinguishable from a genuine CME signal. Therefore, understanding and constraining the behavior of $v_2$ is a prerequisite for any meaningful interpretation of CME observables. Here, our primary objective is to classify events by their geometrical shape, so that events with lower elliptic flow contributions can be selected to identify CME signals better.\\

In this analysis, we used the elliptic flow coefficient $v_2$ measured from the inclusive final-state charged particle distribution, rather than restricting to $v_2$ computed exclusively from identified resonance decay products or jet fragments. Cleanly isolating resonance daughters or jet-associated particles on an event-by-event basis is experimentally intractable, as these particles are kinematically entangled with the bulk medium.
Thus, we cannot identify them without introducing autocorrelations with the particles of interest used to construct $\Delta\gamma$ \cite{Li:2024gdz}. Moreover, the elliptic flow anisotropy of background-contributing sources, $v_{2,\text{res}}$, is related to the elliptic flow coefficient $v_2$. $v_{2,\text{res}}$ is driven by the same underlying collision geometry that governs the bulk collective expansion. As a result, $v_{2,\text{res}} \propto v_2$, which means that selecting events based on the inclusive $v_2$ (or equivalently, the transverse spherocity $S_0$) simultaneously affects the background strength encoded in $\Delta\gamma_{\text{res}}$ (Eq. \ref{eq:deltagammares}).
Figure \ref{fig:v2} presents the $v_2$  of charged particles as a function of transverse momentum  ($p_T$). The $v_2$ values measured for different event classes were selected using transverse spherocity cuts in Pb+Pb collisions at $\sqrt{s_{NN}}=5.02$ TeV. The results are shown for integrated events (red squares) and for events classified as jetty (blue stars) and isotropic (black circles) using several selection criteria: (70\%-30\%), (80\%-20\%), and (90\%-10\%) in Fig. \ref{fig:v2}(a),(b), and (c), respectively. Here the notation (X\%-Y\%) denotes selecting the top X\% of events as isotropic and the bottom Y\% as jetty in the spherocity distribution.\\

A systematic hierarchy in $v_2$ is observed across all $p_T$ bins. The measured $v_2$ is largest in jetty events, intermediate in integrated events, and smallest in isotropic events. This trend is robust and persists for all applied spherocity cuts. However, the magnitude of the difference between jetty and isotropic classes varies with the strictness of the selection. The (90\%;10\%) cut, which isolates the top and bottom 10\% events in the spherocity distribution (ref. Fig.\ref{fig:v2}(c), yields the largest separation in $v_2$ between the two classes. This confirms that transverse spherocity effectively disentangles events with different underlying dynamics. Jetty events are dominated by flow-related correlations from hard processes and jet fragmentation, which enhance the measured $v_2$, while isotropic events represent the collective bulk medium with minimal jet contamination.\\

The relevance of this result for our CME study is twofold. First, it validates spherocity as an effective tool for event-shape engineering, creating distinct environments in which flow-driven backgrounds differ significantly across event classes. Second, and more critically, it provides the necessary baseline for interpreting the charge-dependent $\Delta\gamma$ correlator. Since a major background to $\Delta\gamma$ scales approximately with $v_2$ , the observation that 
$v_2$ is substantially lower in isotropic events directly implies that the flow-related background should be suppressed in this class. Consequently, if a residual $\Delta\gamma$ signal persists or is even enhanced in isotropic events after proper scaling by $v_2$, it would constitute stronger evidence for a contribution from the CME, which our earlier results suggest favors an isotropic topology. In section \ref{sec:resC}, we will investigate the $\Delta\gamma$ correlator within this established framework.

\subsubsection{Resonance decays as a background in spherocity-selected events}

\begin{figure*}[ht]
\begin{minipage}[b]{0.49\linewidth}
    \centering
    \begin{overpic}[scale=0.34]{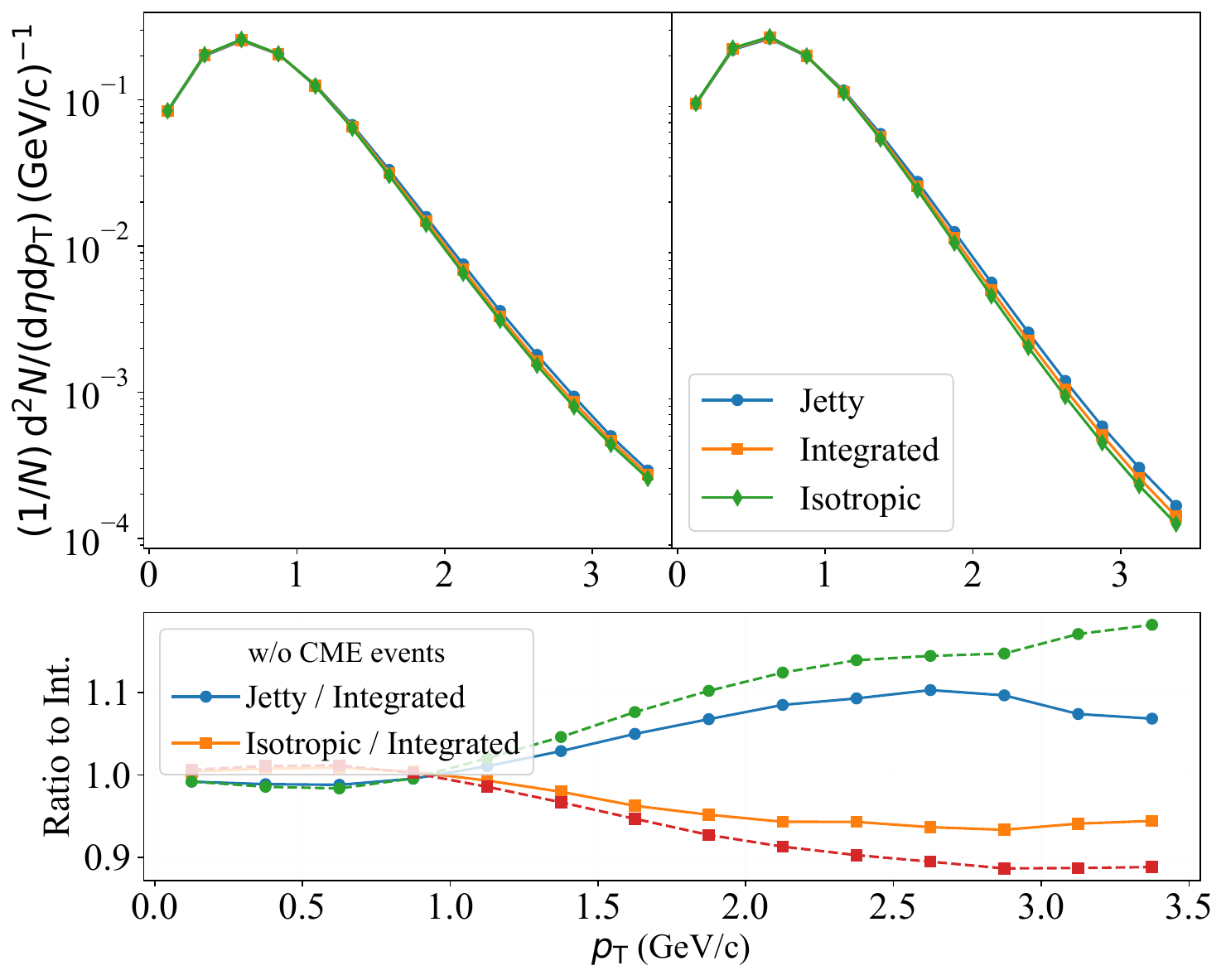}
    \put(82,70){\textbf{(a)}}
    \put(29,74){Spherocity cut}
    \put(32,69){(70\%-30\%)}
    \put(14,54){w/o CME}
    \put(57,54){with CME}
    \end{overpic}
\end{minipage}
\hfill
\begin{minipage}[b]{0.49\linewidth}
    \centering
    \begin{overpic}[scale=0.34]{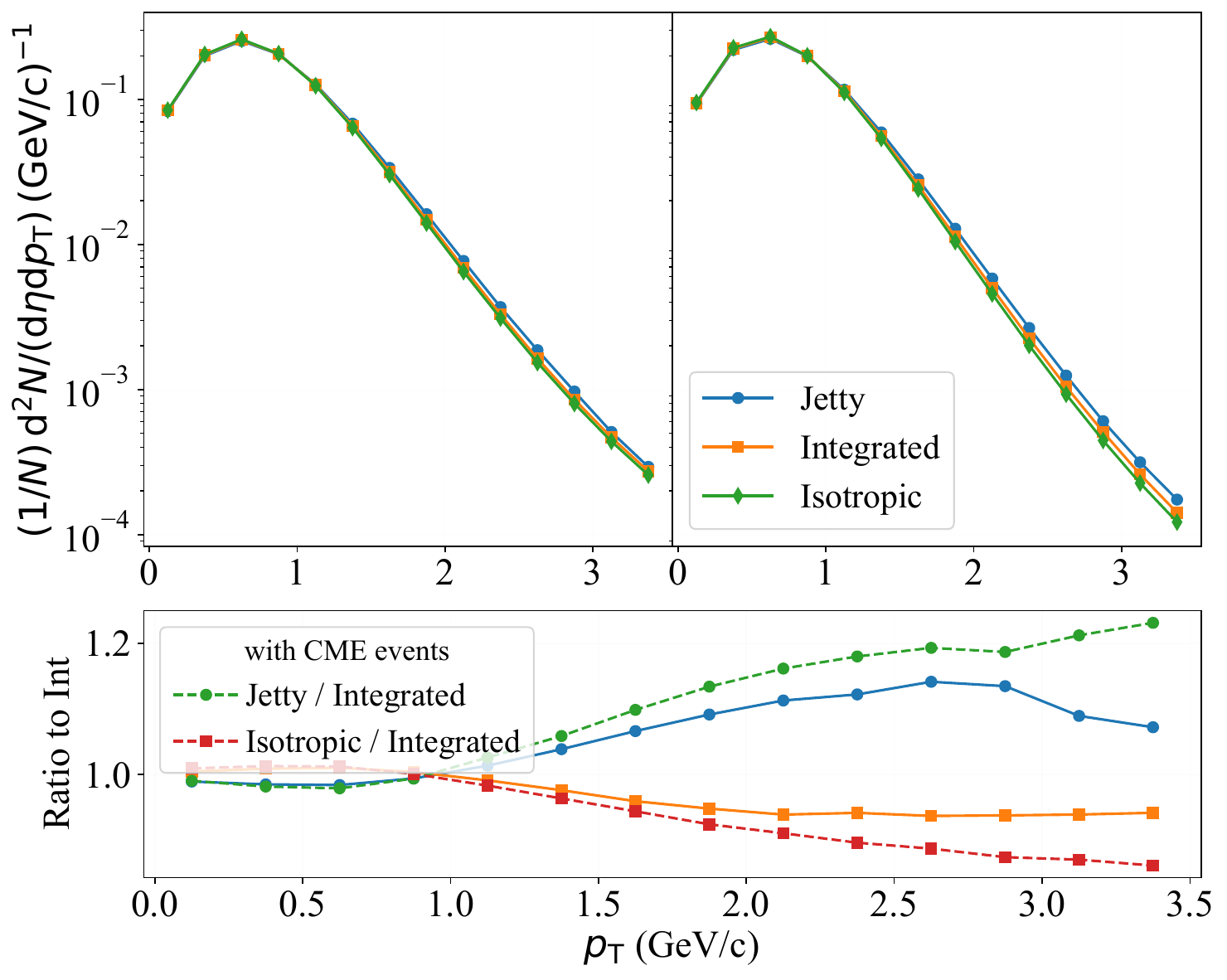}
        \put(82,70){\textbf{(b)}}
        \put(29,74){Spherocity cut}
        \put(32,69){(80\%-20\%)}
        \put(14,54){w/o CME}
        \put(57,54){with CME}
    \end{overpic}

\end{minipage}
\centering
\begin{overpic}[scale=0.36]{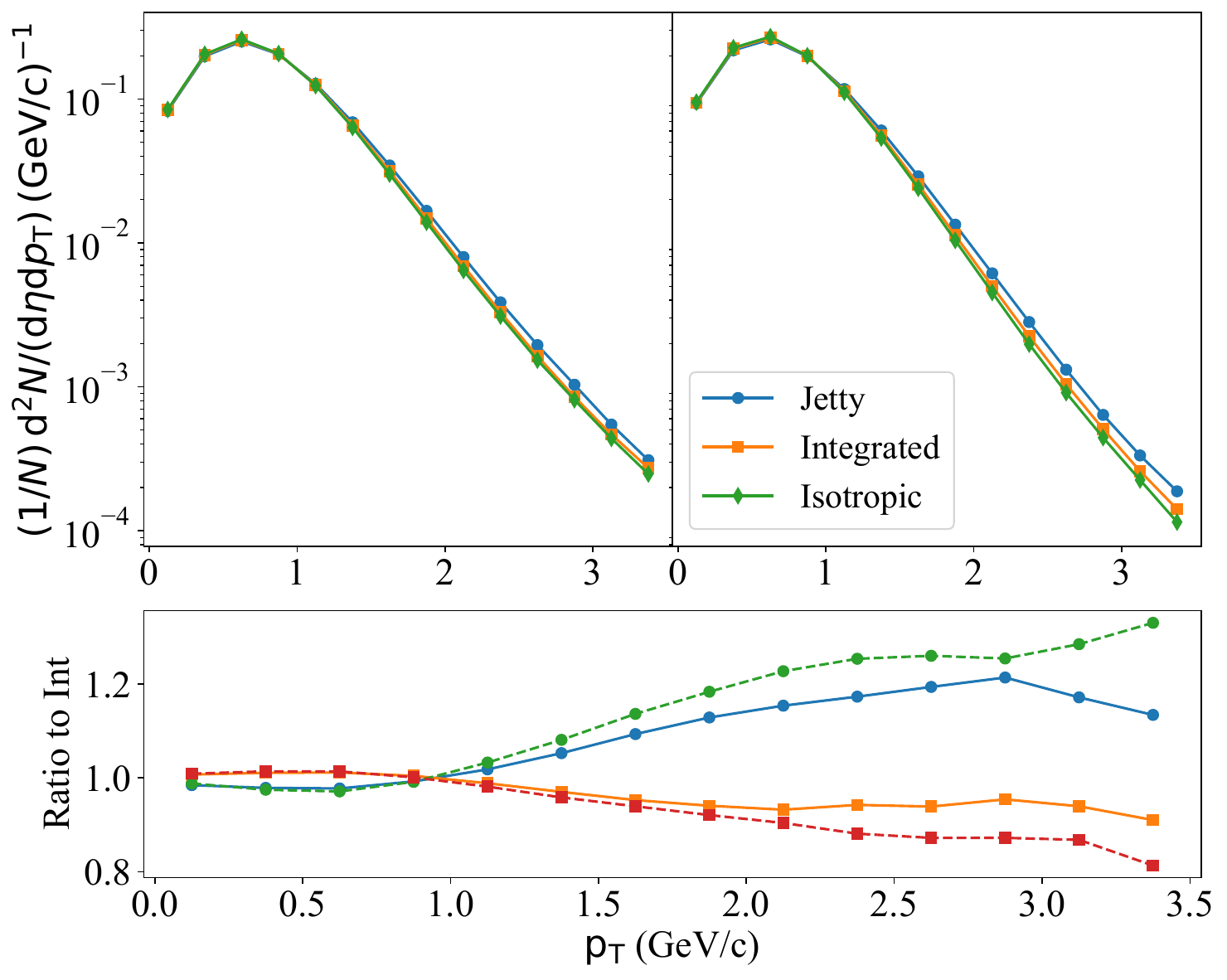}
    \put(82,70){\textbf{(c)}}
    \put(29,74){Spherocity cut}
    \put(32,69){(90\%-10\%)}
    \put(14,44){Pb+Pb (30-50\%)}
    \put(18,39){5.02 TeV}
    \put(14,56){w/o CME}
    \put(57,56){with CME}
\end{overpic}

\caption{Transverse momentum spectra of $K^{*0}$ mesons in Pb+Pb collisions at $\sqrt{s_{NN}} = 5.02$ TeV for three spherocity selections: (a) $70\%$-$30\%$, (b) $80\%$-$20\%$, and (c) $90\%$-$10\%$. The upper left panels show results without CME, and the upper right panels show results with CME. The bottom panels present the ratio of jetty and isotropic yields to the integrated yield.}
\label{fig:res1}
\end{figure*}

\begin{figure*}[ht]
\begin{minipage}[b]{0.49\linewidth}
    \centering
    \begin{overpic}[scale=0.34]{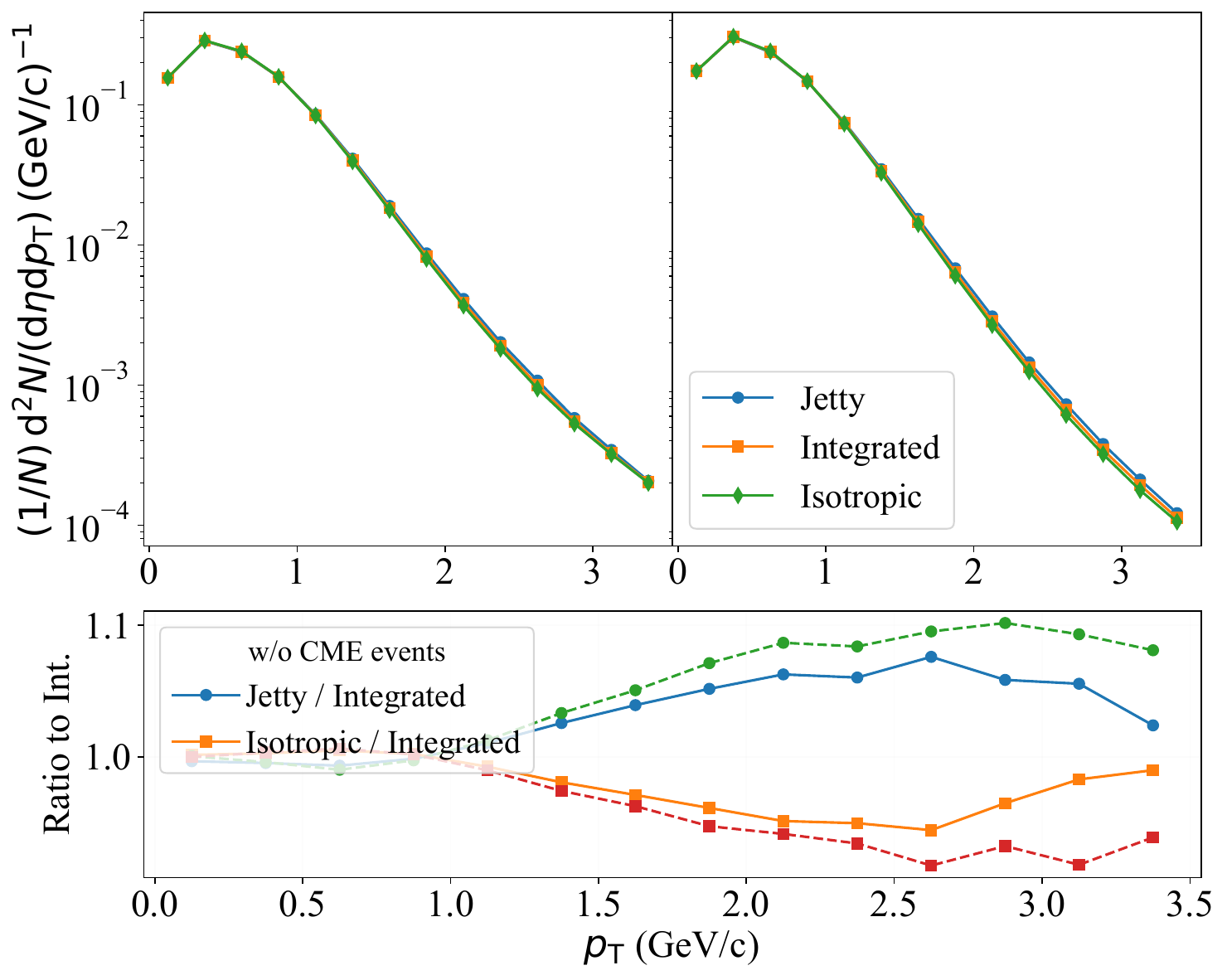}
    \put(82,70){\textbf{(a)}}
    \put(29,74){Spherocity cut}
    \put(32,69){(70\%-30\%)}
    \put(14,54){w/o CME}
    \put(57,54){with CME}
    \end{overpic}
\end{minipage}
\hfill
\begin{minipage}[b]{0.49\linewidth}
    \centering
    \begin{overpic}[scale=0.34]{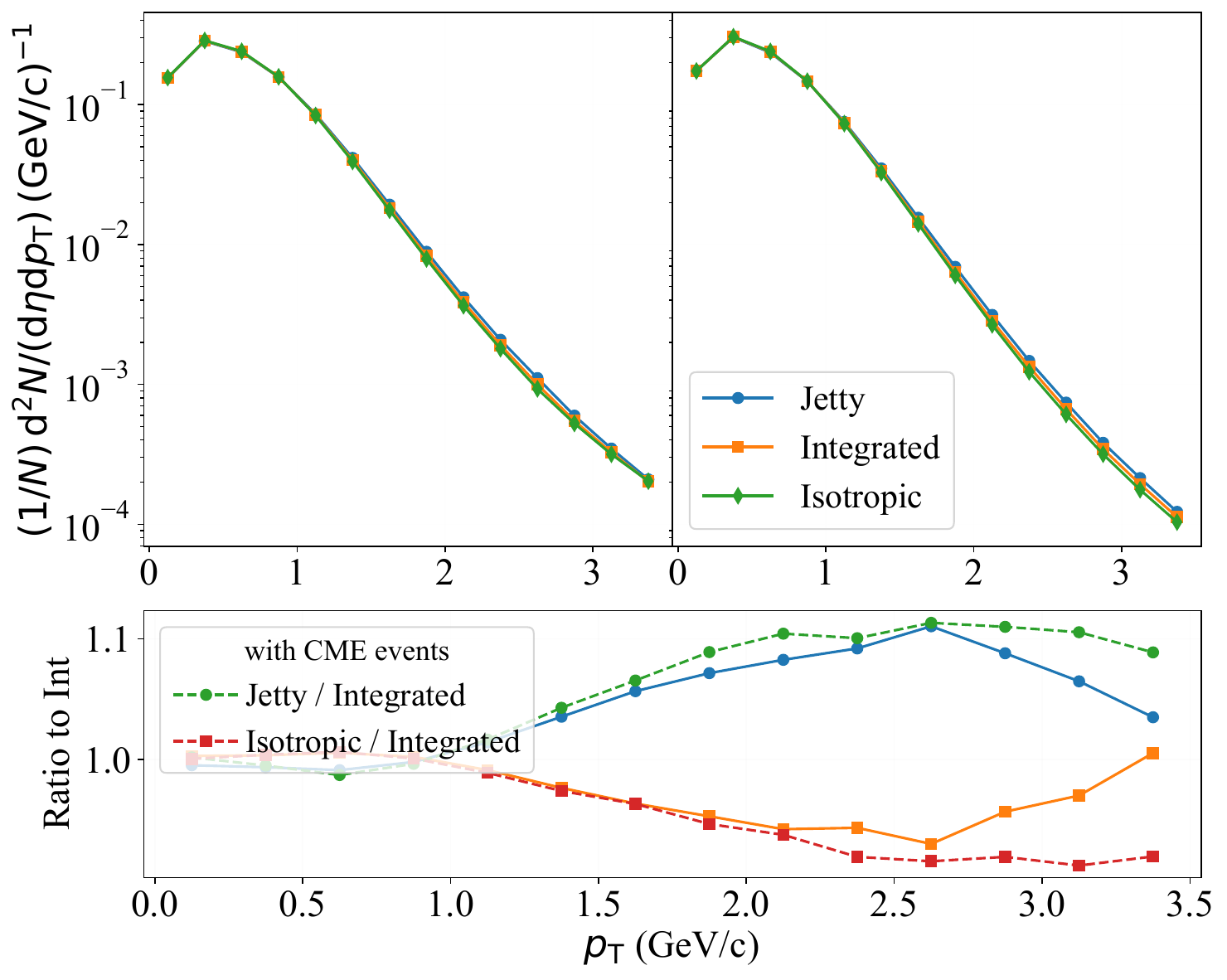}
        \put(82,70){\textbf{(b)}}
        \put(29,74){Spherocity cut}
        \put(32,69){(80\%-20\%)}
        \put(14,54){w/o CME}
        \put(57,54){with CME}
    \end{overpic}
\end{minipage}
\centering
\begin{overpic}[scale=0.36]{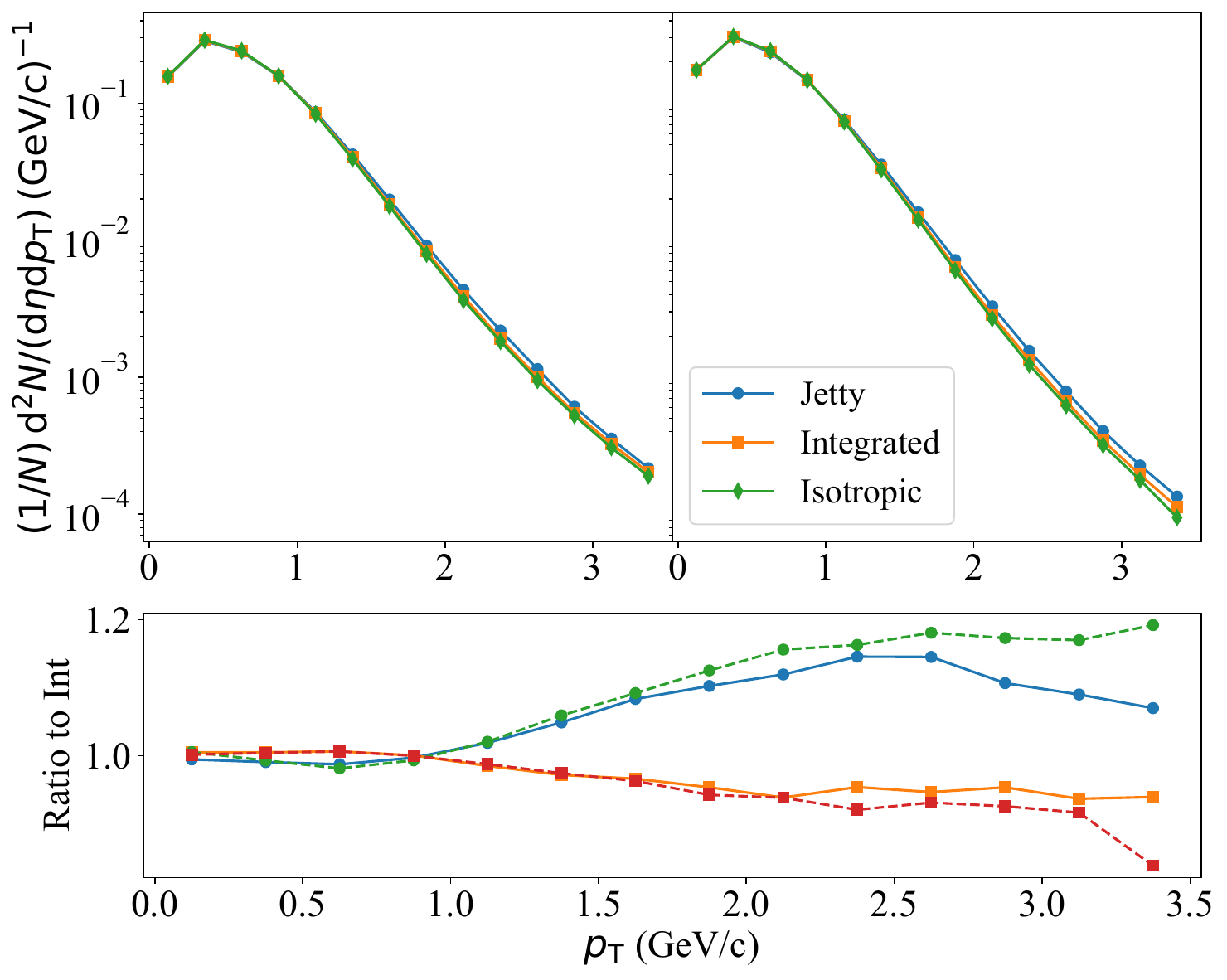}
    \put(82,70){\textbf{(c)}}
    \put(29,74){Spherocity cut}
    \put(32,69){(90\%-10\%)}
    \put(14,44){Pb+Pb (30-50\%)}
    \put(18,39){5.02 TeV}
    \put(14,54){w/o CME}
    \put(57,54){with CME}
\end{overpic}
\caption{Transverse momentum spectra of $\rho^{0}$ mesons in Pb+Pb collisions at $\sqrt{s_{NN}} = 5.02$ TeV for three spherocity selections: (a) $70\%$-$30\%$, (b) $80\%$-$20\%$, and (c) $90\%$-$10\%$. The upper left panels show results without CME, and the upper right panels show results with CME. The bottom panels present the ratio of jetty and isotropic yields to the integrated yield.}
\label{fig:res2}
\end{figure*}


It has been observed that resonance decays associated with elliptic flow can mimic the CME signal \cite{Li:2024pue, Xu:2023elq, Tang:2019pbl}. These decays produce charge-dependent azimuthal correlations between decay products that mimic the charge separation. When a short-lived resonance decays, its daughter particles (often a pair of charged hadrons) are emitted with closely correlated momenta in azimuth and rapidity. This correlation contributes to the measured $\Delta\gamma$ correlator. They introduce a charge-dependent angular correlation that is not related to the reaction plane or topological gluon fields. This background is particularly significant because, like the CME, it is intrinsically charge-dependent. However, its origin is purely hadronic and local. It does not arise from a macroscopic current induced by a magnetic field.\\

We focus on resonances that are abundantly produced in heavy-ion collisions, decay into charged particles, and are found to affect the CME signal the most. In our study, we considered two specific resonances: the $K^{*0}(892)$ and $\rho^0(770)$ mesons. The $K^{*0}$ decays via $K^{*0}(\bar{K^{*0}})\rightarrow K^{+}(K^{-})+\pi^{-}(\pi^{+})$, while the $\rho^0$ decays via $\rho^0\rightarrow\pi^+ + \pi^-$ \cite{Li:2024pue, Tang:2019pbl}.  Both are two-body decays that produce opposite-sign charged particle pairs with strong angular correlations. These two resonances have well-defined, high-branching-ratio decay channels into charged kaons and pions that fall within standard experimental acceptance. Furthermore, they represent different quark content. The $K^{*0}$ carries strangeness ($s\bar{d}$/$d\bar{s}$), while the $\rho^0$ is composed of light quarks ($u\bar{u}-d\bar{d}$). This allows us to probe whether backgrounds from resonance decays depend on quark flavor or mass.\\


We reconstruct resonances using the invariant mass ($m_{\text{inv}}$) distribution of charged hadron pairs. This method is standard in experiments like ALICE and STAR; thus, our model study is comparable to experimental measurements \cite{ALICE:2023bga, ALICE:2017ban}. In the AMPT model, resonance production and decay are naturally part of the hadronization and hadronic rescattering stages \cite{STAR:2008bgi, Tang:2019pbl}. The CME we introduce in AMPT changes the invariant-mass distributions for opposite-sign and same-sign pairs. Previous studies have used AMPT to investigate non-flow contributions from resonance decays \cite{Zhang:2023kin} and to study $K_S$ and $\rho$ decays using the invariant mass method in CME analyses \cite{Zhao:2019hta}.\\

We analyze the model event record directly to identify resonance decay vertices. For the $K^{*0}$ and $\rho^0$, we reconstruct their yields by calculating the invariant mass of all candidate charged daughter pairs ($K^{\pm}\pi^{\pm}$ for $K^{*0}$ and $\pi^{+}\pi^{-}$ for $\rho^0$) within each event. We identify true decay products by matching daughter particles to their parent resonance in the event history. We then study the $p_T$-differential yield and azimuthal correlations of these decay products as a function of event shape, classified by transverse spherocity. The $p_T$ spectra for the reconstructed $K^{*0}$ and $\rho^0$ reveal how resonance decay backgrounds change with event topology. Figure \ref{fig:res1} presents the $p_T$-differential yields of $K^{*0}$ for three spherocity cuts: (a) 70\%–30\%, (b): 80\%–20\%, (c): 90\%–10\%, We present results for the 30–50\% centrality class, with events grouped into jetty, integrated, and isotropic categories. 
The upper panels show the $p_T$ spectra for cases without CME (upper left panels) and with CME (upper right panels) separately. The lower panels show the ratio of yields in each event class to the integrated yield, again separated into the cases without CME and with CME. Figure \ref{fig:res2} presents the corresponding yields for the $\rho^0$ meson.

A clear pattern appears across all spherocity cuts. The resonance yield is highest in jetty events, intermediate in integrated events, and lowest in isotropic events. This holds for both mesons and does not depend on whether the CME is present. The bottom panels of each figure quantify this by showing the ratio of the jetty or isotropic yield to the integrated yield. The jetty-to-integrated ratio is consistently greater than 1, while the isotropic-to-integrated ratio is always less than 1. This confirms that resonance production is significantly stronger in jet-dominated event topologies and suppressed in isotropic events.

The excess in jetty events indicates that the processes defining a jetty topology also favor the production of resonances like $K^{*0}$ and $\rho^0$. These processes originate from hard parton scatterings and the subsequent fragmentation. These resonances can form directly in jet fragmentation or in the enhanced dense medium surrounding a jet trajectory. Thus, the same event characteristic that leads to large non-flow and elliptic flow also amplifies this background source to the CME. The suppression of isotropic events indicates that these events, characterized by uniform, soft momentum flow, have a lower relative abundance of resonance decays. The trend grows stronger with stricter spherocity selection.
As the cut is tightened from (70\%–30\%) to (90\%–10\%), the jetty-to-integrated ratio for both $K^{*0}$ and $\rho^0$ increases from $\sim1.15$ to $\sim1.3$. At the same time, the isotropic-to-integrated ratio decreases from $\sim0.9$ to $\sim0.8$. We also checked for heavier resonances, such as $\phi$ and $\Delta$. For them, the enhancement and suppression are even larger.
This divergence demonstrates that stricter spherocity selection more effectively classifies the event classes. The effect is stronger for heavier particles. This suggests that strange-quark production or the formation of heavier resonances couples more strongly to jet-like activity.

\subsection{Charge correlator in spherocity selected events}\label{sec:resC}
We have seen how backgrounds to the CME signal differ for event selections based on transverse spherocity. Jetty events are background-rich, exhibiting concurrent enhancements in $v_2$ and resonance yields. Any CME-like signal in this class is heavily contaminated. In contrast, isotropic events background-suppressed, showing concurrent reductions in $v_2$, resonance yields. A strict spherocity selection, such as the (90\%–10\%) cut, provides the most effective background suppression by maximizing the difference between jetty and isotropic environments. Selecting isotropic events under such a cut minimizes contributions from resonance decays and related flow-driven correlations, thereby enhancing the relative sensitivity to any remaining CME signal component. In this section, we present a systematic study of the CME-sensitive observable $\Delta\gamma$ and its scaled counterpart $\Delta\gamma/v_2$ as a function of transverse spherocity to quantify how signal sensitivity evolves with event topology.

\subsubsection{Charge Correlators}
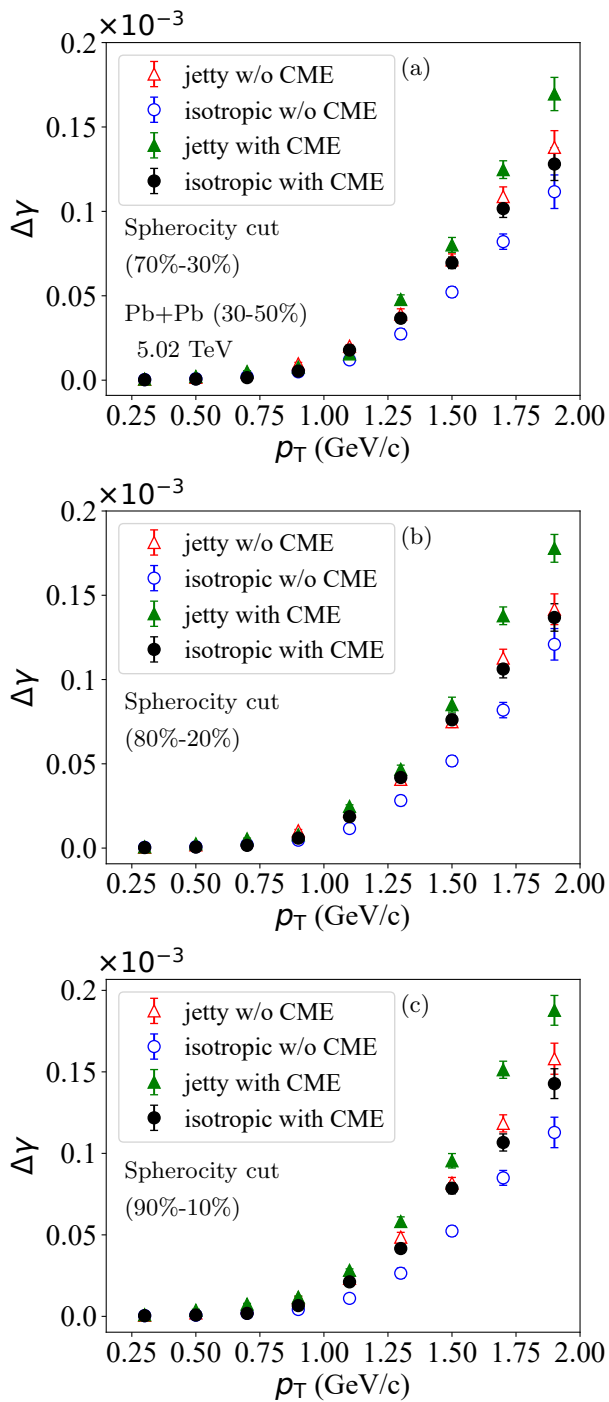
\begin{figure}[h!]
\centering

\begin{overpic}[scale=0.5]{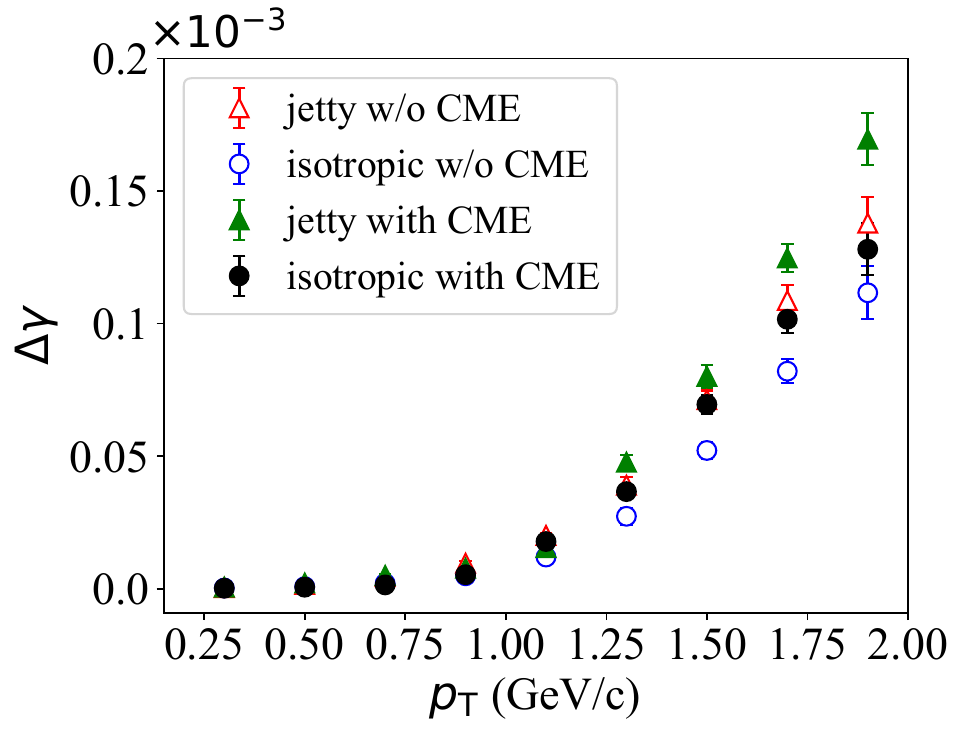}
\put(65,65){(a)}
\put(20,39){Spherocity cut}
\put(20,33){(70\%-30\%)}
\put(20,25){Pb+Pb (30-50\%)}
\put(22,19){5.02 TeV}
\end{overpic}
\begin{overpic}[scale=0.5]{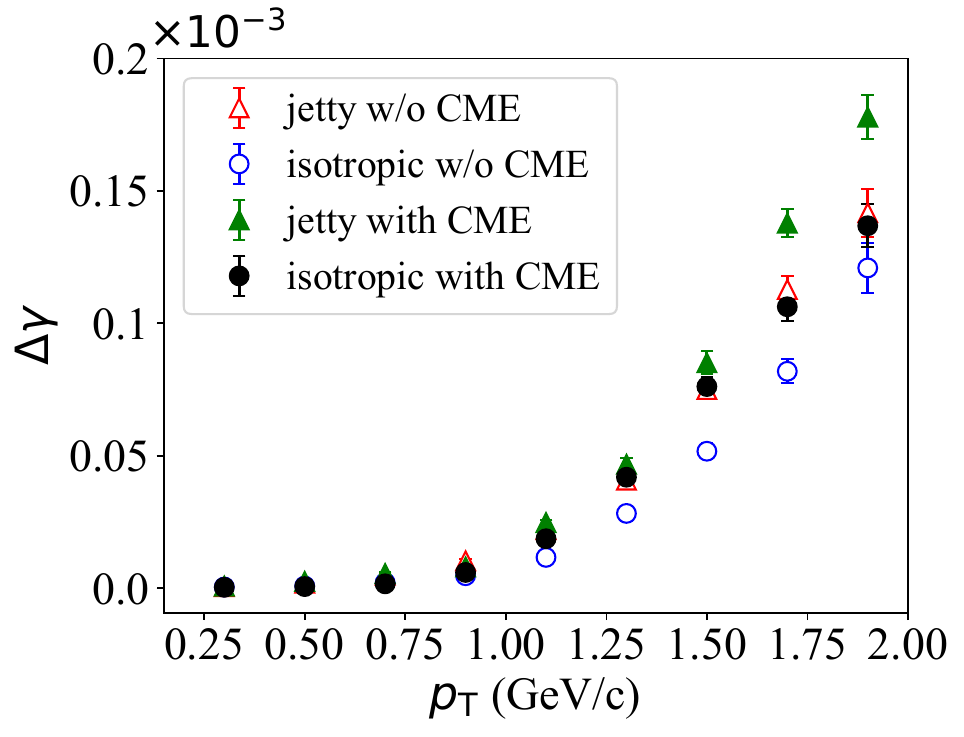}
\put(65,65){(b)}
\put(20,38){Spherocity cut}
\put(20,32){(80\%-20\%)}
\end{overpic}
\begin{overpic}[scale=0.5]{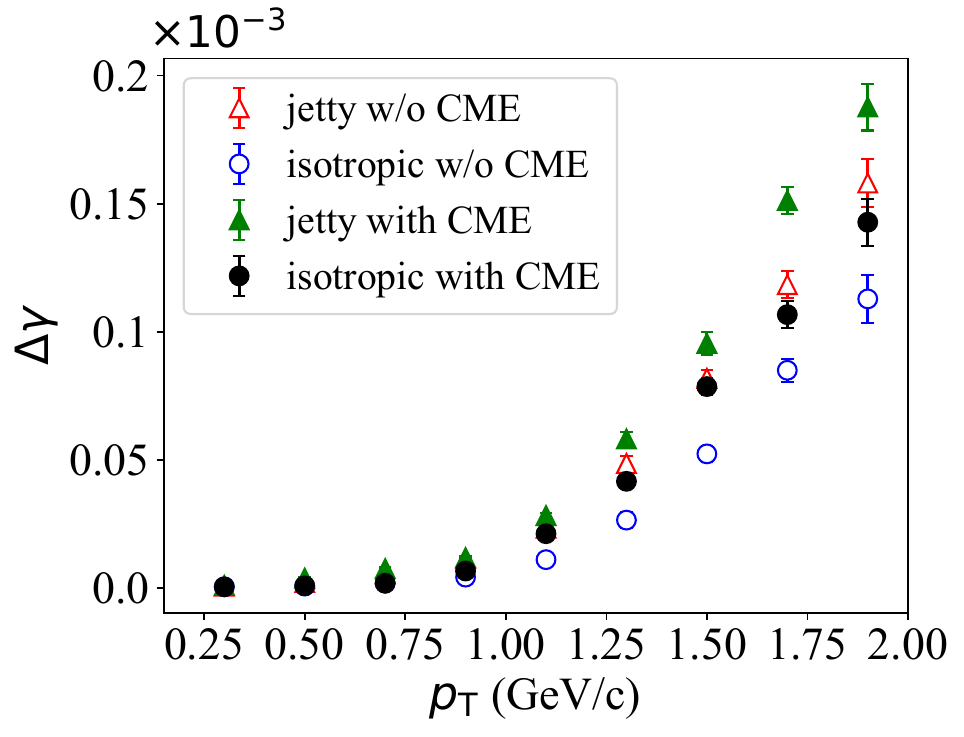}
\put(65,65){(c)}
\put(20,38){Spherocity cut}
\put(20,32){(90\%-10\%)}
\end{overpic}
\caption{Transverse momentum dependence of the charge-dependent correlator $\Delta\gamma$ in Pb–Pb collisions at $\sqrt{s_{NN}}= 5.0$ TeV. The panels correspond to different spherocity selections: (a) 70\%–30\%, (b) 80\%–20\%, and (c) 90\%–10\%. Results are presented for the AMPT model both without and with the CME implementation.}
\label{fig:gamma1}
\end{figure}
Figure \ref{fig:gamma1} presents the charge-dependent correlator $\Delta\gamma$, defined in Eq. \ref{eq:deltagamma}, as a function of transverse momentum $p_T$ for different transverse spherocity selections in Pb+Pb collisions at $\sqrt{s_{NN}}=5.02$ TeV (30-50\% centrality). The figure compares the behavior of $\Delta\gamma$ across different event topologies, e.g., isotropic and jetty events, for three distinct spherocity selection criteria: (a) 70\%–30\%, (b) 80\%–20\%, and (c) 90\%–10\%. Crucially, the results are shown both with (solid markers) and without (open markers) the explicit implementation of the CME in the AMPT model. 

A key observation across all panels is that $\Delta\gamma$ values are consistently higher in jetty events than in isotropic events, for both scenarios with and without the CME implementation. This trend holds across the entire measured $p_T$  range and becomes more pronounced with stricter spherocity selection, as seen in the progression from panel (a) to (c). This trend aligns with the measured hierarchy in elliptic flow $v_2$ (Fig. \ref{fig:v2}), where jetty events also exhibited larger values. This correlation strongly suggests that the dominant contribution to $\Delta\gamma$ in the "w/o CME" case arises from background processes, such as resonance decays and local charge conservation coupled with anisotropic flow, which are inherently enhanced in jetty topologies due to their larger non-flow and flow contributions.\\

The $\Delta \gamma$ signal for isotropic events with CME (black circles) remains remarkably stable across all three spherocity cuts. This consistency indicates that the CME contribution, which our earlier analysis showed pushes events towards isotropy (Sec-\ref{sec:resA}), manifests as a robust, additive component in the most isotropic environments. Its magnitude is largely independent of the exact spherocity threshold once a baseline isotropic class is defined. In contrast, the  $\Delta \gamma$ signal for jetty events with CME (green circles) is systematically higher than its isotropic counterpart. It grows with stricter spherocity selection, as it compounds the genuine CME signal on top of an increasingly large jetty background.\\

The difference between the two classes ($\Delta\gamma_{\text{jetty}}-\Delta\gamma_{\text{isotropic}}$) can be treated as a measure of the background magnitude. Therefore, the isotropic event class provides a stable environment for CME searches. In these events, the background is minimized and stable, so any excess in $\Delta \gamma$ (the consistent offset seen in the black circles vs. blue circles) can be more reliably attributed to a CME-like effect. The increasing separation between jetty and isotropic events in the "w/o CME" scenario with stricter cuts further validates that selecting the most isotropic events (e.g., via the (90\%–10\%) cut) effectively suppresses the background-dominated regime.
The following section will explicitly examine the ratio $\Delta\gamma/v_2$  to quantify this signal-to-background enhancement.



\subsubsection{Scaled correlator and background suppression}

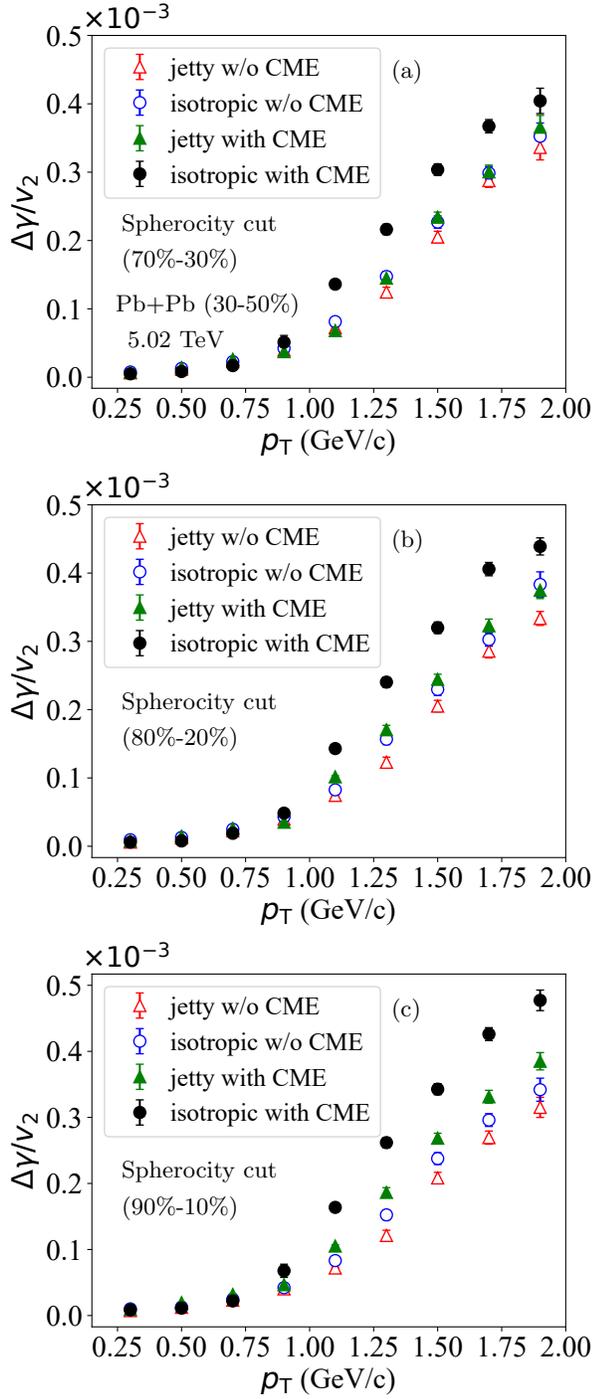
\begin{figure}[h!]
\centering
\begin{overpic}[scale=0.5]{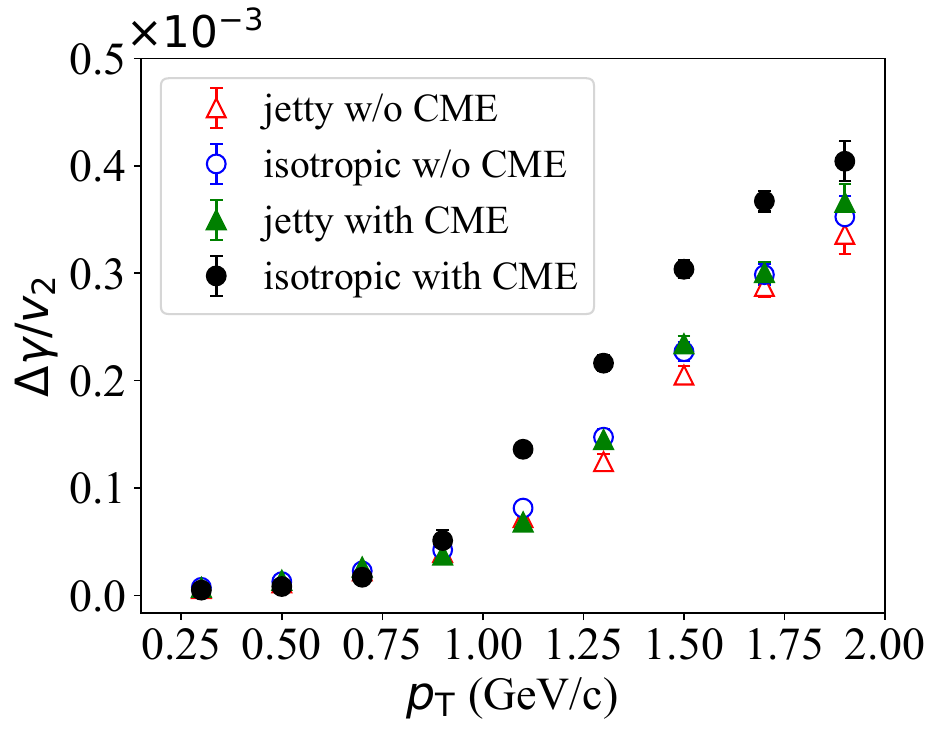}
\put(65,65){(a)}
\put(20,39.5){Spherocity cut}
\put(20,33.5){(70\%-30\%)}
\put(19,26){Pb+Pb (30-50\%)}
\put(21,20){5.02 TeV}
\end{overpic}
\begin{overpic}[scale=0.5]{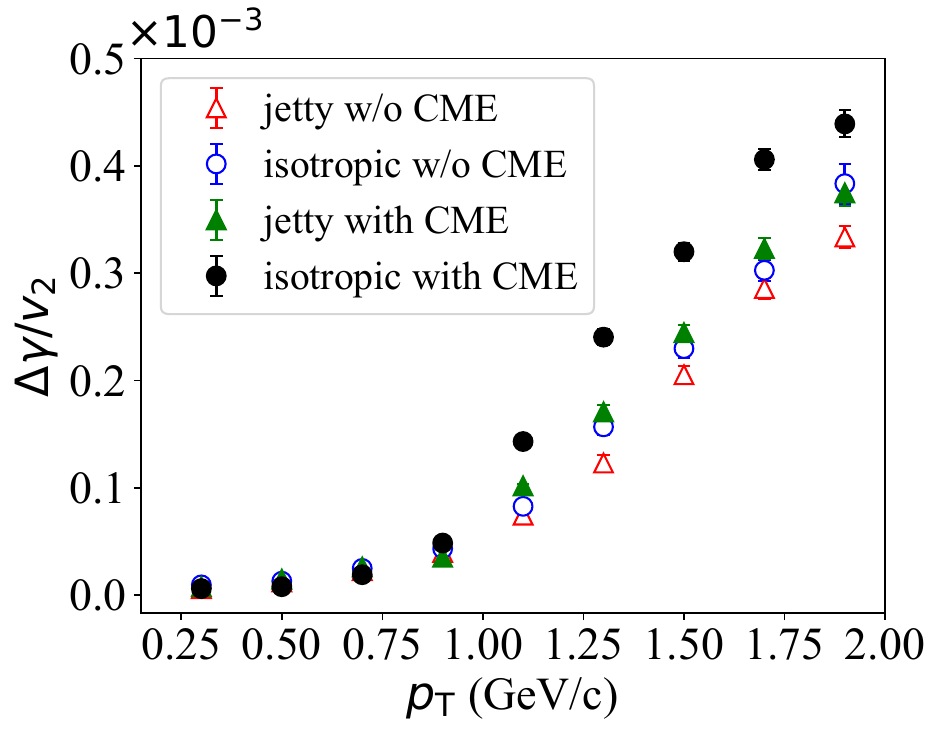}
\put(65,65){(b)}
\put(20,38){Spherocity cut}
\put(20,32){(80\%-20\%)}
\end{overpic}

\begin{overpic}[scale=0.5]{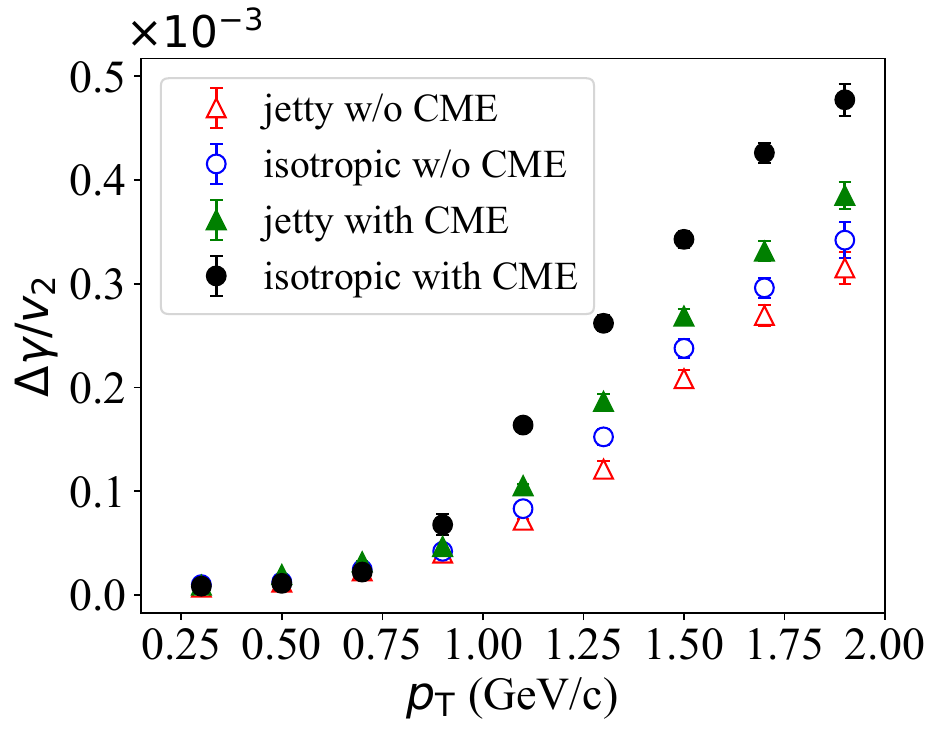}
\put(65,65){(c)}
\put(20,38){Spherocity cut}
\put(20,32){(90\%-10\%)}
\end{overpic}
\caption{Transverse momentum dependence of the scaled charge-dependent correlator $\Delta\gamma/v_2$ in Pb–Pb collisions at $\sqrt{s_{NN}}= 5.0$ TeV. The panels correspond to different spherocity selections: (a) 70\%–30\%, (b) 80\%–20\%, and (c) 90\%–10\%. Results are presented for the AMPT model both without and with the CME implementation.}

\label{fig:gc_v21}
\end{figure}
To directly probe the signal-to-background sensitivity for the chiral magnetic effect, we examine the scaled correlator $\Delta\gamma/v_2$. This ratio is a crucial experimental measurement since the dominant background processes, namely local charge conservation and resonance decays coupled with collective flow, are expected to scale linearly with the $v_2$. Therefore, 
$\Delta\gamma/v_2$ helps to isolate any signal component that does not follow this scaling, such as a potential CME contribution.\\

Figure \ref{fig:gc_v21} presents $\Delta\gamma/v_2$ as a function of $p_T$ for the same event classes and spherocity selections as shown in Fig. \ref{fig:gamma1}. The results reveal a consistent pattern that confirms the significance of spherocity-based event engineering. In all panels, the scaled correlator for isotropic events with CME (black circles) is significantly higher than for any other classes, including jetty events with CME. Furthermore, the behavior under different spherocity cuts follows a systematic trend. For the loose ($70\%–30\%$) cut, the enhancement of $\Delta\gamma/v_2$ in isotropic CME events is present but modest. The values for jetty events (with and without CME) and isotropic events without CME all cluster at a lower, similar level. As the spherocity selection becomes stricter, the separation increases. The 
$\Delta\gamma/v_2$ value for isotropic CME events rises substantially, while the values for the other three classes remain low and closely grouped. The $(90\%–10\%)$ cut provides the most pronounced enhancement, offering the largest distinction between the potential signal and the background-dominated regimes.\\

This behavior is fully consistent with our earlier results and provides a coherent explanation for the utility of spherocity as an event shape classifier in CME searches. Our analysis showed that isotropic events have a smaller $v_2$ (Fig. \ref{fig:v2}), indicating reduced flow-driven backgrounds. The CME implementation contributes a stable, additive component to $\Delta\gamma$ in isotropic events (Fig. \ref{fig:gamma1}), which is largely independent of the spherocity cut.
The combination of these two effects leads to a larger $\Delta\gamma/v_2$ ratio in isotropic CME events. Stricter spherocity cuts further purify the isotropic class, minimizing $v_2$ even more without diminishing the CME's contribution to $\Delta\gamma$, thereby maximizing the scaled observable.\\

\begin{figure}[h!]
\begin{overpic}[scale=0.5]{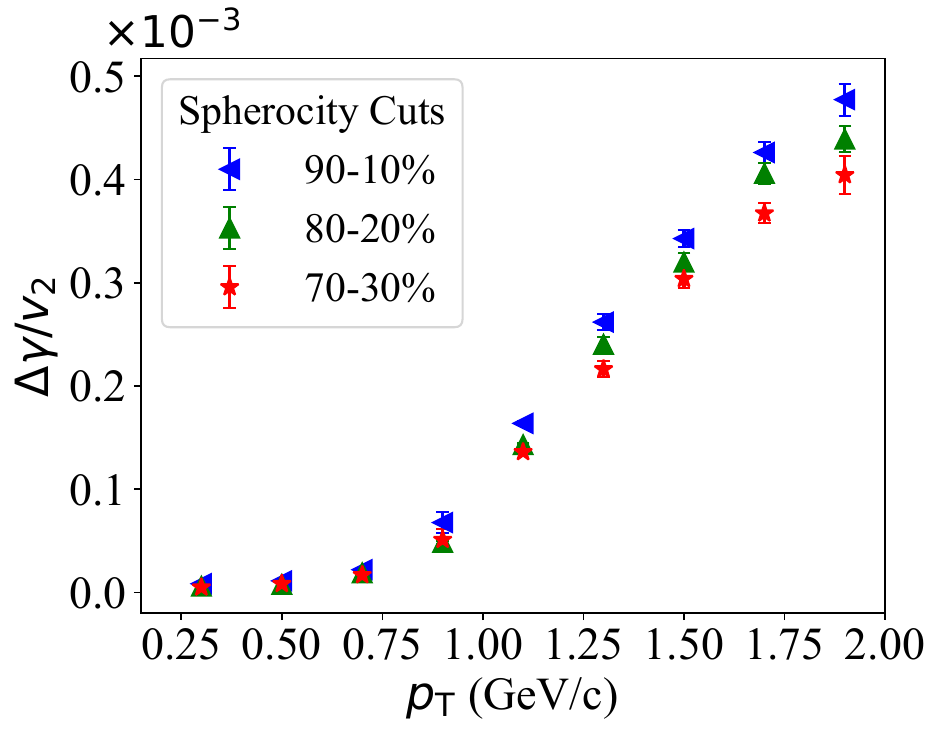}
\put(18,38){Isotropic events}
\put(20,32){with CME}
\put(52,66){Pb+Pb (30-50\%)}
\put(54,60){5.02 TeV}
\end{overpic}

\caption{Transverse momentum dependence of the scaled charge-dependent correlator $\Delta\gamma / v_{2}$ in Pb+Pb collisions at $\sqrt{s_{NN}} = 5.02$ TeV. The panels correspond to different isotropic spherocity selections: (a) 70\%–30\%, (b) 80\%–20\%, and (c) 90\%–10\%. These results are obtained from the AMPT model with the CME implementation.} 
\label{fig:gc_v2}
\end{figure}
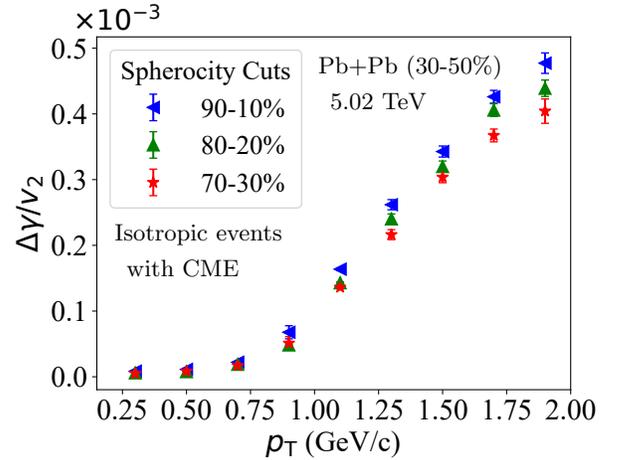

Figure \ref{fig:gc_v2} reinforces this conclusion by isolating the comparison for events with the CME implemented scaled charge-dependent correlator $\Delta\gamma / v_2$ for different transverse spherocity selections within the isotropic event class. The systematic enhancement of  $\Delta\gamma / v_{2}$ in isotropic events, which increases with the strictness of the spherocity cut, provides a compelling model-based strategy for CME detection. Therefore, measuring $\Delta\gamma / v_{2}$ in events classified by transverse spherocity offers a promising approach for experimental searches. A significant, persistent excess of this ratio in real data, specifically within the isotropic event class, would constitute robust evidence consistent with the chiral magnetic effect.\\

The results from both $\Delta\gamma$ and $\Delta\gamma / v_2$  lead to two central conclusions. First, transverse spherocity successfully categorizes events into background-rich (jetty) and background-suppressed (isotropic) regimes. Second, the signal characteristics of the implemented CME make a stable additive component to $\Delta\gamma$ that does not scale with $v_2$, which are most clearly observable in the isotropic event class, particularly under the strictest (90\%–10\%) spherocity selection. This identifies a concrete analysis strategy for experimental CME searches. Applying a tight spherocity cut to select isotropic events can simultaneously suppress flow-driven and resonance-decay backgrounds, thereby isolating a cleaner charge-separation signal for further scrutiny. 

\section{Conclusion}
\label{sec:conclusion}
In this paper, we presented the first comprehensive study of the Chiral Magnetic Effect (CME) using event-shape engineering via transverse spherocity in Pb+Pb collisions at $\sqrt{s_{NN}}$=5.02 TeV within the AMPT model. By implementing a charge-separation mechanism to simulate the CME and classifying events into jetty and isotropic topologies, we systematically analyzed how the CME signal and its dominant backgrounds behave across different event classes.\\


We found that implementing CME shifts the transverse spherocity distribution toward more isotropic events, showing that spherocity is directly sensitive to CME-induced dynamics. This alone establishes spherocity as a meaningful probe of CME-related physics, beyond just a geometric classifier.
Across all observables, we observed a clear and consistent pattern. Jetty events are characterized by larger elliptic flow and higher resonance decay yields, both of which are major sources of background to the CME signal. In contrast, isotropic events show suppression in all these background quantities, creating a cleaner environment for CME searches. The scaled correlator $\Delta\gamma/v_2$, which is the most experimentally relevant quantity for isolating the CME, is significantly enhanced in isotropic events when the CME is present. This enhancement becomes stronger with stricter spherocity selection, becoming most pronounced for the (90\%–10\%) cut. This result arises from the combination of a stable CME contribution to $\Delta\gamma$ and significantly reduced $v_2$ in isotropic events, maximizing the signal-to-background ratio.\\

The importance of this study for the heavy-ion community is manifold. It introduces transverse spherocity as a new and effective discriminant for event-shape engineering in CME searches, complementary to traditional flow-vector-based methods. It provides a coherent, model-based framework that demonstrates that background contributions are not uniform but are strongly correlated with global event topology. This understanding moves the field beyond integrated measurements and toward a differential analysis strategy through event topology that can separate signal-like from background-like event environments.\\

For experimental collaborations at the LHC and RHIC, our results suggest a concrete path forward. To improve the sensitivity of CME searches, we propose implementing transverse spherocity as an event classifier in experimental datasets, applying a strict isotropic event selection and measuring $\Delta\gamma/v_2$ differentially, and comparing the results with those from jetty events and spherocity-integrated events.
A persistent and growing enhancement of $\Delta\gamma/v_2$ in the isotropic event class, one that strengthens with stricter spherocity cuts, would be a strong indicator of a genuine CME signal, well-separated from conventional flow-driven and resonance-decay backgrounds. The next step is to apply this methodology to experimental data to search for the predicted enhancement. Extending the analysis to other collision systems such as p+Pb and Xe+Xe, and across different collision energies, would further test whether spherocity-based signal enhancement is a universal feature.

\section*{Acknowledgments}

The authors thank Soma Sanyal for insightful discussions. A.S. is grateful to the participants of Hot Quarks 2025 for useful conversations. This work is funded by the IoE research grant No.UH/RITE/PHY/SS/IoE-RC522020/01. S.D. is partially supported by the IoE grant.

\bibliography{ref.bib}

\end{document}